\documentclass[11pt, a4paper, logo, copyright]{googledeepmind}

\pdftrailerid{redacted}

\makeatletter
\renewcommand\bibentry[1]{\nocite{#1}{\frenchspacing\@nameuse{BR@r@#1\@extra@b@citeb}}}
\makeatother

\usepackage{kantlipsum, lipsum}
\usepackage{dsfont}
\usepackage{gdm-colors}
\usepackage[utf8]{inputenc}   
\usepackage{newunicodechar}   
\usepackage{amssymb}          
\usepackage{wasysym,marvosym}
 
\usepackage[authoryear, sort&compress, round]{natbib}

\usepackage{bbding}
\usepackage[utf8]{inputenc} 
\usepackage[T1]{fontenc}    
\usepackage{hyperref}       
\usepackage{url}            
\usepackage{booktabs}       
\usepackage{nicefrac}       
\usepackage{microtype}      
\usepackage{amsmath}
\usepackage{graphicx}
\usepackage{tablefootnote}
\usepackage{multicol}
\usepackage[nameinlink]{cleveref}
\usepackage{bbm}
\usepackage{multirow}
\usepackage{soul}
\usepackage{float}
\usepackage{wrapfig}
\usepackage{blindtext}
\usepackage{tablefootnote}
\usepackage{amsfonts}
\usepackage[flushleft]{threeparttable}
\usepackage{colortbl}
\usepackage{mathtools,amssymb}
\usepackage{bm}
\usepackage{makecell}
\usepackage{caption}
\usepackage{capt-of}
\usepackage{array}
\usepackage{calc}      
\usepackage{caption}   
\usepackage{subcaption}  
\usepackage{xcolor,colortbl}
\usepackage[bottom]{footmisc}

\usepackage{xspace}

\newcommand{\eat}[1]{}

\crefformat{section}{\S#2#1#3}

\graphicspath{{figures/}}

\title{OneRec-V2 Technical Report}

\renewcommand{\today}{}

\author{\large OneRec Team}

\begin{abstract}
 
Recent breakthroughs in generative AI have fundamentally transformed recommender systems by enabling end-to-end generation. OneRec, an industrial-scale generative recommendation framework, reformulates recommendation as an autoregressive generation task, allowing for direct optimization of the final objective and achieving high Model FLOPs Utilization (MFU). 
While OneRec-V1 has shown significant empirical success in real-world deployment, two critical challenges hinder its scalability and performance: (1) \textit{Inefficient computational allocation in encoder-decoder architecture}, where 97.66\% of resources are consumed by sequence encoding context encoding rather than generation, which limits model scalability; and (2) \textit{limitations in reinforcement learning that relies solely on reward models}, including inefficient sampling and potential reward hacking due to proxy reward signals. To address these challenges, we propose \textbf{OneRec-V2}, featuring:  

1. \textbf{Lazy Decoder-Only Architecture}: A streamlined, decoder-only design that eliminates encoder bottlenecks and simplifies cross-attention, reducing total computation by \textbf{94\%} and training resources by \textbf{90\%} (see Figure \ref{fig:flops_vs_loss} right).
This efficiency enables the successful scaling of the model to 8B parameters and, notably, the convergence loss closely follows the empirical scaling law. As the model scales, we observe a smooth and predictable decrease in loss consistent with the scaling law fit (see Figure~\ref{fig:flops_vs_loss} left, and Figure~\ref{fig:dense_params_vs_loss}).

2. \textbf{Preference Alignment with Real-World User Interactions}: A user feedback-driven framework incorporating (i) Duration-Aware Reward Shaping to mitigate video duration bias and (ii) Adaptive Ratio Clipping to stabilize policy optimization, effectively leveraging real-world feedback to better align with user preferences and resulting in a significant increase in App Stay Time.

Extensive A/B tests on Kuaishou/Kuaishou Lite demonstrate the effectiveness of OneRec-V2, improving \textbf{App Stay Time by 0.467\%/0.741\%} while balancing multi-objective recommendations without seesaw effects. This work advances generative recommendation scalability and alignment with real-world feedback, representing a step forward in the development of end-to-end recommender systems. 

\end{abstract}

\begin{document}
\maketitle

\begin{figure}[h!]
    \centering
    \begin{minipage}{0.56\textwidth}
        \centering
        \includegraphics[width=\textwidth]{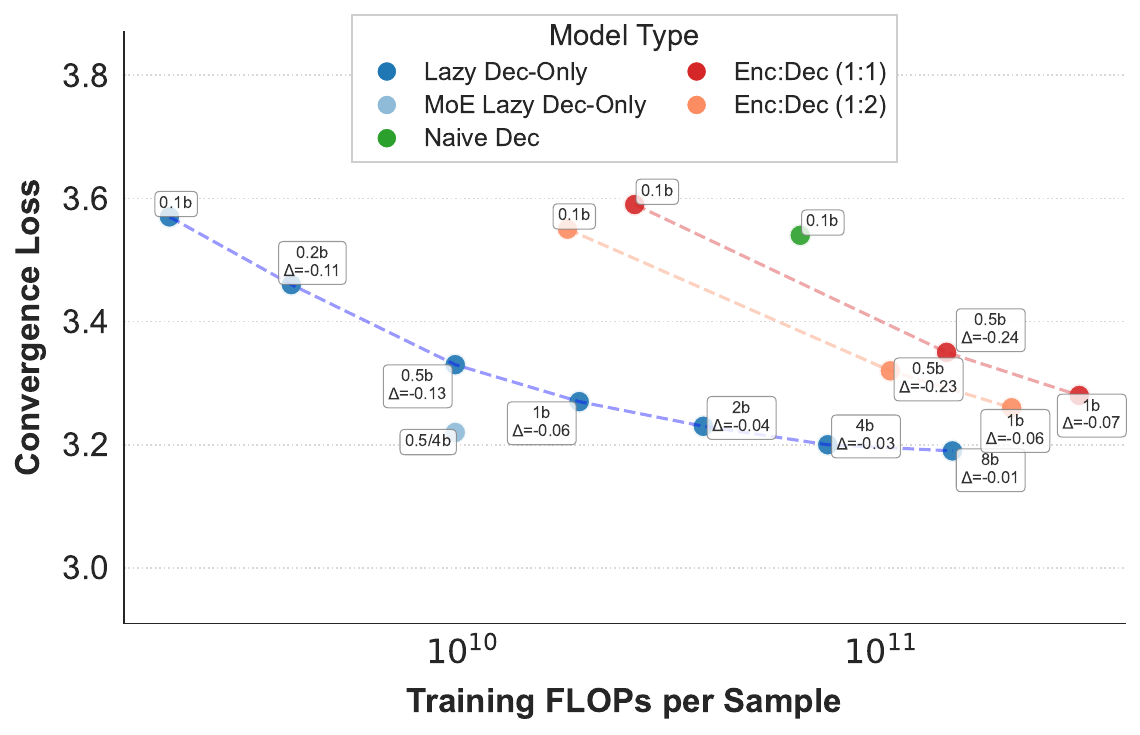}
        \label{fig:flops_vs_loss_main}
    \end{minipage}
    \hfill
    \begin{minipage}{0.28\textwidth}
        \centering
        \includegraphics[width=\textwidth]{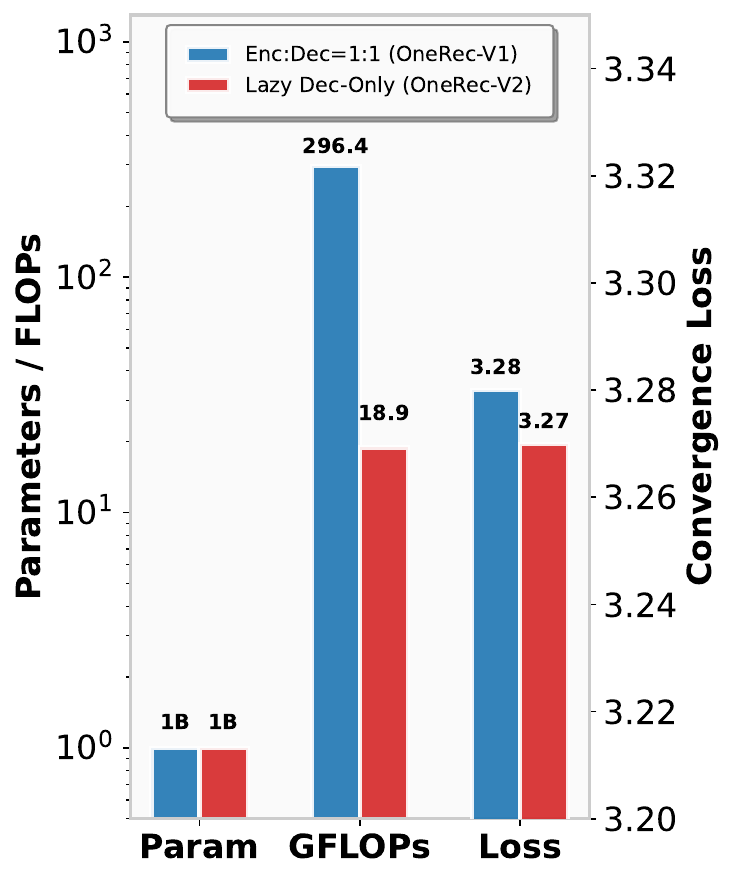}
        \label{fig:onerec_comparison}
    \end{minipage}
    \caption{Left: Scaling curves for various model architectures from 0.1B to 8B parameters, among which Lazy Decoder-only models demonstrate best scaling efficiency. Right: OneRec-V1 v.s. OneRec-V2 at 1B parameters.}
    \label{fig:flops_vs_loss}
\end{figure}

\newpage
\setcounter{tocdepth}{2} 

\tableofcontents 

\newpage
\section{Introduction}
\label{sections:1_introduction}
Generative AI has catalyzed a paradigm shift across numerous domains~\citep{achiam2023gpt, guo2025deepseek, yang2025qwen3}. 
While traditional cascaded recommendation architectures have undergone continuous evolution, they remain constrained by fundamental bottlenecks: the inherent multi-stage design leads to fragmented computational resources and misaligned optimization objectives. Generative recommendation transforms this paradigm by reframing recommendation as an end-to-end sequence generation problem~\citep{cui2022m6, feng2022recommender, rajput2023recommender, zhai2024actions, han2025mtgr, zhou2025onerec, HLLM, kong2025generative, yang2025sparse, badrinath2025pinrec}. This unified approach enables direct optimization of the final objective, achieves high Model FLOPs Utilization
(MFU), and fosters closer integration between recommender systems and large foundation model communities.

While OneRec-V1~\citep{zhou2025onerec} has demonstrated considerable success in industrial deployment, there remain opportunities to further unlock its scalability and performance:

(1) \textbf{Inefficient computational allocation in encoder-decoder architecture}. OneRec-V1 employs an encoder-decoder framework where user historical interaction sequences are processed through encoder, and then utilized by the decoder through cross-attention. Although OneRec-V1's decoder contains more parameters than the encoder, the computational load is predominantly concentrated on the encoder, as it processes extensive user interaction sequences while the decoder's input is significantly shorter. As illustrated in Section~\ref{sec:design}, with context length $512$ in OneRec-V1, the context encoding consumes 97.66\% of total FLOPs, while the target item generation of decoder is merely 2.34\%. This disproportionate allocation presents scalability challenges, as the majority of computational budget is dedicated to sequence encoding rather than the critical generation process where recommendation decisions are formulated. Under equivalent computational budgets, this imbalanced resource distribution may limit the model's potential to scale effectively to larger architectures.

(2) \textbf{limitations in reinforcement learning that relies solely on reward models.} 
Although OneRec-V1 has demonstrated the effectiveness of reward-model-based reinforcement learning for policy optimization, this approach faces two inherent challenges. First, there is \textbf{limited sampling efficiency}, as methods relying on reward models require additional computational resources for online generation and scoring. This restricts sampling to a small subset of users to approximate global behavior. Second, there is \textbf{potential reward hacking}, where the policy learns to exploit specific patterns or biases in the reward model that do not translate to actual improvements. Integrating real user feedback to address these inherent issues could better align the policy with user preferences and lead to improved outcomes. \textit{In addition, OneRec's deployment at a significant scale provides a critical opportunity for self-improvement through policy optimization within a continuous feedback loop.}

\begin{figure}[t]
    \centering
    \includegraphics[width=\textwidth]{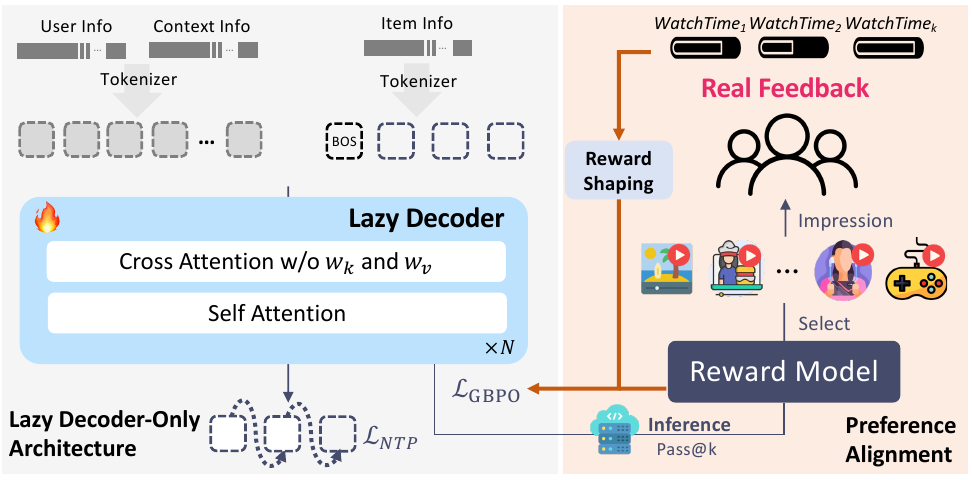}
    \caption{The overall architecture and post-training framework of OneRec-V2. The left panel illustrates the Lazy Decoder-Only Architecture, The right panel depicts the post-training preference alignment process}.
    \label{fig:framework}
\end{figure}

In this work, we introduce OneRec-V2, which addresses these fundamental limitations through a lazy decoder architecture and preference alignment with real-world user interactions. As shown in Figure~\ref{fig:framework}, our key contributions are: 
\begin{enumerate}
    \item \textbf{Lazy Decoder-Only Architecture}. We propose a streamlined decoder-only architecture that eliminates the computational bottleneck of traditional encoder-decoder designs. By removing the encoder component and simplifying cross-attention mechanisms (eliminating K/V projection layers), our lazy decoder achieves a 94\% reduction in computational requirements and 90\% reduction in actual training resources while supporting 16$\times$ larger model parameters (from 0.5B to 8B) under equivalent computational budgets. As shown in Figure~\ref{fig:flops_vs_loss}, this architecture not only makes decoder-only transformers practical and efficient for industrial-scale recommendation systems, but also exhibits strong scaling capabilities: the convergence loss closely follows the theoretical scaling law proposed by Hoffmann et al. (2022)~\citep{hoffmann2022training} across a wide range of model sizes. This provides both empirical and theoretical guidance for the future development of large generative recommendation models.
    \item \textbf{Preference Alignment with Real-World User Interactions}. We introduce a comprehensive post-training framework that directly leverages real-world user feedback signals to address the fundamental challenges of reward modeling in generative recommender systems. (i) \textit{Duration-Aware Reward Shaping}, which mitigates the inherent bias in raw watch time signals by accounting for video length variations, ensuring that reward signals accurately reflect content quality rather than merely duration; and (ii) \textit{Adaptive Ratio Clipping}, which effectively reduces training variance while preserving convergence guarantees in the policy optimization process. Our experiments demonstrate significant gains in \textit{APP Stay Time}. Notably, we observe amplified online performance when incorporating traffic distribution patterns from OneRec's own recommendations, suggesting improved alignment between model optimization and real-world user behavior distributions. 
    
\end{enumerate}

Extensive online A/B testing on Kuaishou/Kuaishou Lite APP with 400 million daily active users demonstrates that OneRec-V2 achieves significant improvements compared to OneRec-V1, delivering improvements of 0.467\% and 0.741\% in App Stay Time, while effectively balancing multiple recommendation objectives without seesaw effects.

In the remainder of this paper, we first elaborate on the OneRec-V2 architecture and empirical results of pre-training (Section~\ref{Architecture}). Next, we present post-training method (Section~\ref{Post-training}), followed by a comprehensive evaluation through online A/B testing (Section~\ref{online_ab_test}). Finally, we conclude this work with a discussion of existing limitations and propose potential directions for future research (Section~\ref{Conclusion}). 

\section{Lazy Decoder-Only Architecture}
\label{Architecture}
In this section, we present the lazy decoder-based architecture. 
Section~\ref{sec:design} elaborates the evolutionary path and thinking of OneRec model architecture.
In Section~\ref{sec:overall_arch}, our lazy decoder-only architecture for OneRec-V2 is presented, which achieves lower generation task loss while significantly reducing both computational complexity and memory consumption.
Finally, comprehensive empirical results across validating the superiority of our lazy decoder-only design, and exploring the scaling laws of generative recommender systems are elaborated in Section~\ref{sec:emp_comparison}.

\begin{figure}[t]
    \centering
    \includegraphics[width=1.0\textwidth]{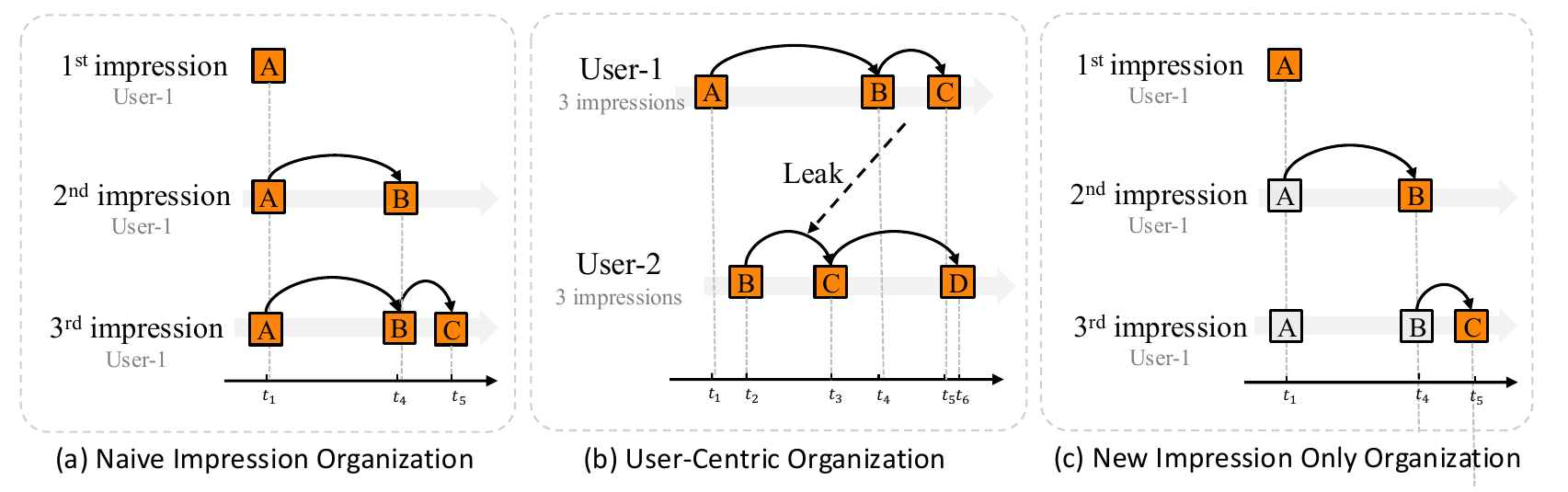}
    \caption{\textbf{Naive Impression Organization}: The pattern A$\rightarrow$B is redundantly trained across multiple impressions. \textbf{User-Centric Organization}: When training on User-2's data at time $t_3$, the model has already learned the pattern B$\rightarrow$C from User-1's future interactions at $t_4$. \textbf{New Impression Only Organization}: It trains only on the newest impression.}
    \label{fig:data_org}
\end{figure}

\subsection{Design Principles}
\label{sec:design}
The autoregressive models have emerged as the dominant paradigm in modern Natural Language Processing, powering state-of-the-art Large Language Models (LLMs) such as GPT~\citep{radford2019language, brown2020language} and LLaMA~\citep{touvron2023llama, touvron2023llama2}. 
They demonstrate remarkable scalability~\citep{kaplan2020scaling, hoffmann2022training}, with their success stemming from elegant simplicity: a unified architecture that processes sequences autoregressively.
Combined with massive-scale pretraining capabilities~\citep{devlin2019bert, raffel2020exploring},  transformer based autoregressive models have become the de facto standard for generative AI applications.

To adapt these architectures to recommender systems, the first step is to construct the \textit{\textbf{doc}} for autoregressive training. Conventionally, the training sample of the recommender system is organized in chronological impression. However, redundancy arises when combined with the standard Next Token Prediction objective, as illustrated in Figure~\ref{fig:data_org}.a.
One way to avoid the redundancy is using user-centric organization, where each training sample encompasses a user's complete interaction history, as illustrated in Figure~\ref{fig:data_org}.b.
However, it carries potential risks of temporal data leakage~\citep{ji2023critical} and popularity bias. Numerous studies~\citep{gharahighehi2021fair,gangwar2021adaptive,zhu2021popularity,huang2022different,klimashevskaia2308survey} have been conducted to mitigate these issues. 

\begin{table}[t]
\centering
\caption{Proportion of computation dedicated to loss-relevant target decoding, calculated for models with 1B parameters. Here the context indicates the user feature tokens Not directly participating in the loss calculation.}
\label{tab:computation_proportion}
\begin{tabular}{lccccc}
\toprule
Context Length N & 512 & 3000 \\
\midrule
\textbf{Encoder-Decoder (0.5B:0.5B)} & &  \\
Total Computation (GFLOPs) & 346 & 1988 \\
Context Encoding  (GFLOPs) & 338 & 1980 \\
Target Decoding  (GFLOPs) & 8.1 & 8.1 \\
\textbf{Target Proportion} & \textbf{2.34\%} & \textbf{0.41\%} \\
\midrule
\textbf{Naive Decoder-Only (1B)} & & & & & \\
Total Computation (GFLOPs) & 632 & 3618 \\
Context Encoding  (GFLOPs) & 614 & 3600 \\
Target Decoding (GFLOPs) & 18 & 18 \\
\textbf{Target Proportion} & \textbf{2.85\%} & \textbf{0.49\%}  \\
\midrule
\textbf{Lazy Decoder-Only (1B)} & & \\
Total Computation (GFLOPs) & 18 & 18\\
\textbf{Target Proportion} & $\mathbf{\approx100\%}$ & $\mathbf{\approx100\%}$ \\
\bottomrule
\end{tabular}
\end{table}

To address above problems, we propose to organize data chronologically but applies the training loss \textit{exclusively} to the newest impressed item, as illustrated in Figure~\ref{fig:data_org}.c, where items in gray are excluded in next token prediction. Since the former and newest impressed items work in different ways, we chose Encoder-Decoder architecture in the previous OneRec-V1 \cite{zhou2025onerec}. 
As shown in Table~\ref{tab:computation_proportion}, we conduct a preliminary analysis of the computation details. The computations can be categorized into two distinct classes, context encoding and target decoding.

DEFINITION 1. \textbf{Context Encoding}

The computational operations that process and transform the user context features, specifically encompassing: (i) the context transformation operations performed in the encoder, and (ii) the context projection operations in the cross-attention of the decoder.

DEFINITION 2. \textbf{Target Decoding}

The computational operations that process and transform \textbf{semantic tokens of the target item} in the decoder, specifically encompassing: (i) the self-attention that captures dependencies among semantic tokens, (ii) the feed-forward network (FFN) that applies non-linear transformations, and (iii) the query and output transformations in the cross-attention.

According to Table~\ref{tab:computation_proportion}, Encoder-Decoder save nearly half computations with identical number of parameters comparing to classic Decoder-Only architecture.
However, both architectures still suffer from computational inefficiency: \textit{the majority of computations are allocated to tokens that do not directly contribute to loss computation}.  
For a typical context length of $N=512$ (OneRec-V1), less than 3\% of total FLOPs contribute to loss computation, which becomes increasingly negligible as context length grows. 
Detailed computation analysis is provided in Appendix~\ref{appendix:computation_ana}. 
To \textbf{concentrate computations exclusively on semantic tokens of the target item}, thereby enabling efficient scaling to larger models, we proposed Lazy Decoder-Only Architecture.

\begin{figure}[t]
    \centering
    \includegraphics[width=1.0\textwidth]{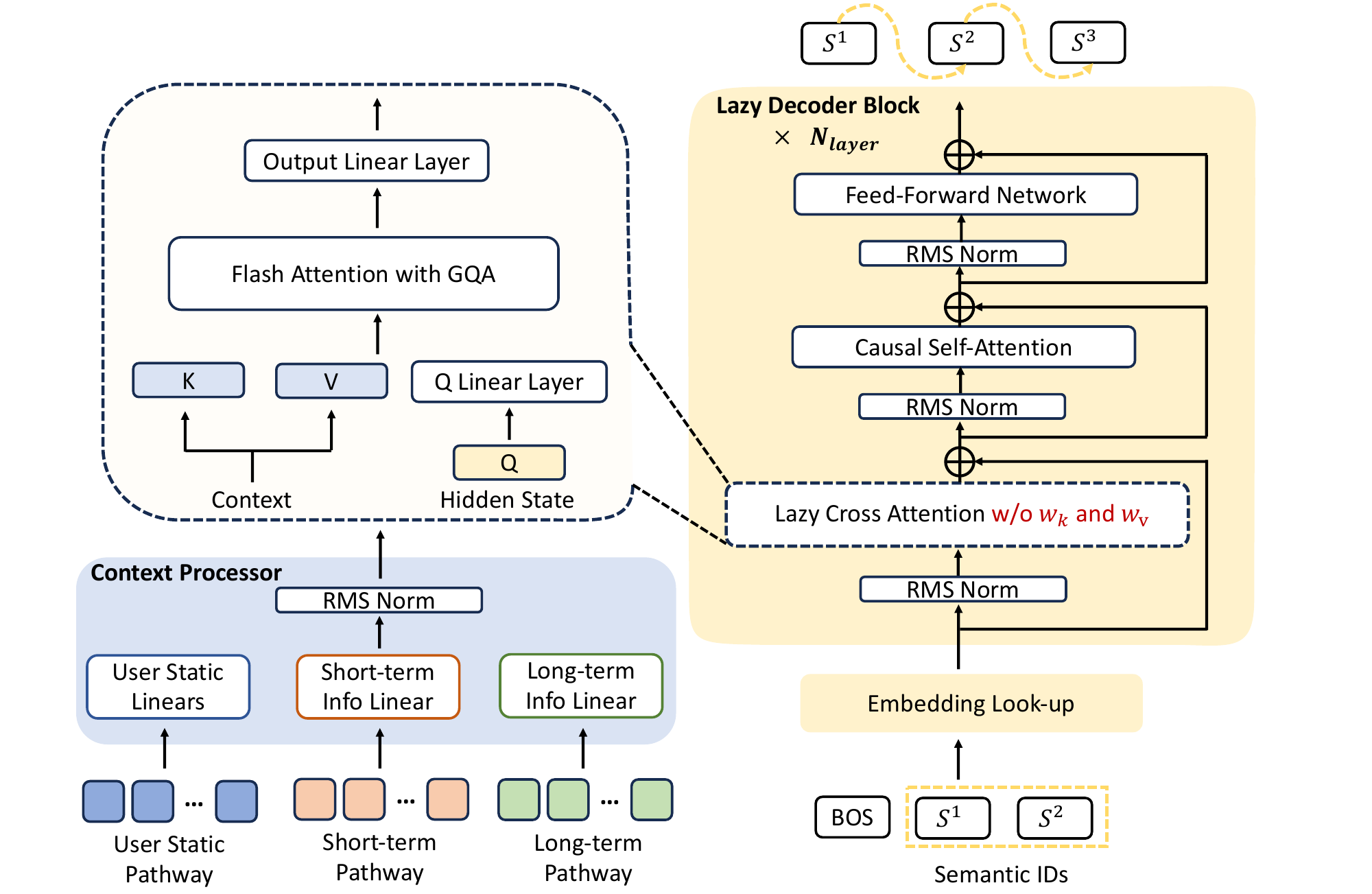}
    \caption{Architecture of the proposed lazy decoder-only generative recommender.
    The Context Processor transforms heterogeneous user feature pathways into unified context representations, which are then normalized to produce layer-shared key-value pairs for cross-attention. 
    The Lazy Decoder processes BOS token and tokenized semantic IDs of the target item through stacked transformer blocks. 
    Each block comprises: (1) lazy cross-attention without key-value projections, enabling Grouped Query Attention (GQA); (2) causal self-attention; and (3) a feed-forward network. 
    The final representations are projected to predict semantic IDs for next-item recommendation.}
    \label{fig:lazy_decoder}
\end{figure}

\subsection{Overall Architecture}
\label{sec:overall_arch}
In this section, we present our novel architecture, illustrated in Figure~\ref{fig:lazy_decoder}, which fundamentally reimagines the design of generative recommenders through two key innovations. 

\textbf{First}, we propose a \emph{lazy decoder-only} architecture that departs from both traditional encoder-decoder models and naive decoder-only approaches. 
Our design treats context as static conditioning information accessed solely through cross-attention, eliminating redundant computation while preserving the model's ability to capture complex user-item interactions.

\textbf{Second}, we introduce an extremely efficient lazy cross-attention mechanism without key-value projections. 
Combined with Grouped Query Attention (GQA)~\citep{ainslie2023gqa}, this design dramatically reduces memory footprint, enabling efficient processing of extensive user histories. 


\subsubsection{Context Processor}
To effectively integrate heterogeneous and multi-modal user behavioral signals, we design a unified module termed the \texttt{Context Processor}, enabling seamless integration with downstream attention-based decoder blocks.





Specifically, heterogeneous inputs such as user profile and behavior are concatenated as an unified sequence, namely context.
Every item in context is processed to identical dimension:
\begin{equation}
d_{\text{context}} =  S_{\text{kv}}\cdot L_{\text{kv}} \cdot G_{\text{kv}} \cdot  d_{\text{head}},
\end{equation}
where $d_{\text{head}}$ denotes the attention head dimension, $G_{\text{kv}}$ the number of key-value head groups, $S_{\text{kv}}$ the key-value split coefficient, and $L_{\text{kv}}$ the number of key-value layers. 



The context representation is transformed into layer-specific key-value pairs for the attention mechanism. 
We partition the context tensor along the feature dimension to generate $L_{\text{kv}}$ sets of key-value pairs:

\begin{equation}
\text{Context} = [\mathbf{C}_0, \mathbf{C}_1, \ldots, \mathbf{C}_{S_{\text{kv}} \cdot L_{\text{kv}} - 1}],
\end{equation}
where $\mathbf{C}_{S_{\text{kv}} \cdot L_{\text{kv}} - 1} \in \mathbb{R}^{G_{\text{kv}} \cdot  d_{\text{head}}}$. Here the sequential dimension is ignored for simplicity.

For each layer $l \in \{0, 1, \ldots, L_{\text{kv}}-1\}$, we compute the normalized key-value pairs:
\begin{equation}
\mathbf{k}_l = \text{RMSNorm}_{k,l}(\mathbf{C}_{l \cdot S_{\text{kv}}}),
\end{equation}

\begin{equation}
\mathbf{v}_l = \begin{cases}
\text{RMSNorm}_{v,l}(\mathbf{C}_{l \cdot S_{\text{kv}} + 1}), & \text{if } S_{\text{kv}} = 2 \text{ (separated key-value)} \\
\mathbf{k}_l, & \text{if } S_{\text{kv}} = 1 \text{ (shared representation)}.
\end{cases}
\end{equation}

The final output of the Context Processor is $\{(\mathbf{k}_0, \mathbf{v}_0), \ldots, (\mathbf{k}_{L_{\text{kv}}-1}, \mathbf{v}_{L_{\text{kv}}-1})\}$.

\subsubsection{Lazy Decoder Block}
\paragraph{Tokenizer}
For each target item, we employ a semantic tokenizer that generates \textbf{3} semantic IDs capturing the item's multi-faceted characteristics as in Onerec-V1~\cite{zhou2025onerec}. 
During training, we utilize the first \textbf{2} IDs and prepend a beginning-of-sequence (BOS) token to form the input sequence. 
These token indices are then mapped through  embedding tables to obtain the initial hidden representation: 
\begin{equation}
\mathbf{h}^{(0)} = \text{Embed}([\text{BOS}, s^1, s^2]) \in \mathbb{R}^{3 \times d_{\text{model}}}.
\end{equation}

\paragraph{Block Structure}
The lazy decoder comprises $N_{\text{layer}}$ stacked transformer blocks, each incorporating three primary components: cross-attention, self-attention, and feed-forward modules. For the $l$-th layer, the transformation is defined as:
\begin{align}
\mathbf{h}_{\text{cross}}^{(l)} &= \mathbf{h}^{(l-1)} + \text{CrossAttn}\left(\text{RMSNorm}(\mathbf{h}^{(l-1)}), \mathbf{k}_{l_{\text{kv}}}, \mathbf{v}_{l_{\text{kv}}}\right), \\
\mathbf{h}_{\text{self}}^{(l)} &= \mathbf{h}_{\text{cross}}^{(l)} + \text{SelfAttn}\left(\text{RMSNorm}(\mathbf{h}_{\text{cross}}^{(l)})\right), \\
\mathbf{h}^{(l)} &= \mathbf{h}_{\text{self}}^{(l)} + \text{FFN}^{(l)}\left(\text{RMSNorm}(\mathbf{h}_{\text{self}}^{(l)})\right),
\end{align}
where RMSNorm denotes root mean square layer normalization for training stability.

To enhance model capacity while maintaining computational efficiency, we adopt a hybrid architecture where dense feed-forward networks in deeper layers are replaced with Mixture-of-Experts (MoE) modules.
Following DeepSeek-V3~\citep{liu2024deepseek}, we employ an auxiliary-loss-free load balancing strategy that ensures efficient expert utilization. 

\paragraph{Lazy Cross-Attention: KV-Sharing}
To promote parameter and computational efficiency, multiple lazy decoder blocks share the same set of key-value pairs derived from the context processor. For the current layer $l$, we determine the corresponding key-value index:
\begin{equation}
l_{\text{kv}} = \left\lfloor \frac{l \cdot L_{\text{kv}}}{N_{\text{layer}}} \right\rfloor,
\end{equation}
where $N_{\text{layer}}$ is the total number of lazy decoder blocks. 
This design ensures that every consecutive blocks share the same contextual representations $(\mathbf{k}_{l_{\text{kv}}}, \mathbf{v}_{l_{\text{kv}}})$, where $\mathbf{k}_{l_{\text{kv}}}, \mathbf{v}_{l_{\text{kv}}} \in \mathbb{R}^{(N_s + T_{\text{short}} + T_{\text{long}}) \times G_{\text{kv}} \times d_{\text{head}}}$.

We further enhance parameter efficiency by employing a unified key-value representation, where $\mathbf{v}_{l} = \mathbf{k}_{l}$ for all layers, leveraging the observation that tied key-value projections can maintain comparable performance while  reducing the model's memory footprint.

\paragraph{Lazy Cross-Attention: Grouped Query Attention}
While the query projection maintains $H_q = d_{\text{model}} / d_{\text{head}}$ attention heads, the key-value pairs utilize only $G_{\text{kv}}$ head groups, where typically $G_{\text{kv}} < H_q$. This design significantly reduces both the memory footprint of context representations and the memory access requirements during attention computation, enabling efficient scaling to longer contexts and larger batch sizes.

\paragraph{Output Layer}
The final hidden representation from the last decoder block undergoes position-specific RMSNorm and Linear layer to generate predictions for each semantic ID. 
During training, we optimize the model to maximize the likelihood of the semantic IDs of the target item $[s^1, s^2, s^3]$. 

\subsection{Empirical Results}
\label{sec:emp_comparison}
To validate the effectiveness of lazy decoder-only architecture, we conduct comprehensive empirical evaluations across multiple dimensions. 
We systematically compare our approach against classic architectures, investigate the impact of key architectural innovations, and explore scaling properties for dense and sparse model variants. 
All experiments are conducted using streaming training on impression data from Kuaishou spanning August 10-14, 2025, with the same sampling ratio and a consistent global batch size.
Unless otherwise specified, we set $L_{\text{kv}} = 1$, $S_{\text{kv}} = 1$, $d_{\text{head}} = d_\text{model} / N_\text{head}$, $G_{\text{kv}} = N_\text{head}$ and $(N_s+T_{\text{short}}+T_{\text{long}})\approx 512$.
For online deployment, we employ a 1B parameter model and expand the long-term user behavior sequence length to $(N_s+T_{\text{short}}+T_{\text{long}})\approx 3000$.

\subsubsection{Architecture Comparison}
\begin{table}[t]
\centering
\caption{Comparison of different architectures across model scales. Naive Decoder-Only experiments at 0.5B and 1B scales were not conducted due to computational resource limitations. The number of activations is calculated under the batch size of 512.}
\label{tab:arch_comparison}
\begin{tabular}{lcrrc}
\toprule
\textbf{Architecture} & \textbf{Total Parameters}\tablefootnote{Approximate 0.1B,0.5B and 1B, according to specific model configurations.} & \textbf{GFLOPs} & \textbf{Activations} & \textbf{Convergence Loss} \\
\midrule
Enc:Dec=1:1 & 0.1B & 25.64 & 4.21B & 3.59 \\
Enc:Dec=1:2 & 0.1B & 17.72 & 2.92B & 3.55 \\
Naive Dec-Only & 0.1B & 63.78 & 7.52B & 3.54 \\
Lazy Dec-Only & 0.1B & 1.98 & 0.31B & 3.57 \\
\midrule
Enc:Dec=1:1 & 0.5B & 142.73 & 10.79B & 3.35 \\
Enc:Dec=1:2 & 0.5B & 104.73 & 7.94B & 3.32 \\
Naive Dec-Only & 0.5B & 317.68 & 19.28B & * \\
Lazy Dec-Only & 0.5B & 9.55 & 0.77B & 3.33 \\
\midrule
Enc:Dec=1:1 & 1B & 296.36 & 17.63B & 3.28 \\
Enc:Dec=1:2 & 1B & 204.21 & 12.20B & 3.26 \\
Naive Dec-Only & 1B & 634.83 & 31.53B & * \\
Lazy Dec-Only & 1B & 18.89 & 1.24B & 3.27 \\
\bottomrule
\end{tabular}
\begin{flushleft}
\footnotesize      
Note: FLOPs and activations in this table are calculated based on specific model configurations, which are more precise  compared to the approximate estimates presented in Table~\ref{tab:computation_proportion}.
\end{flushleft}
\end{table}

\begin{figure}[h]
    \centering
    \includegraphics[width=1.\textwidth]{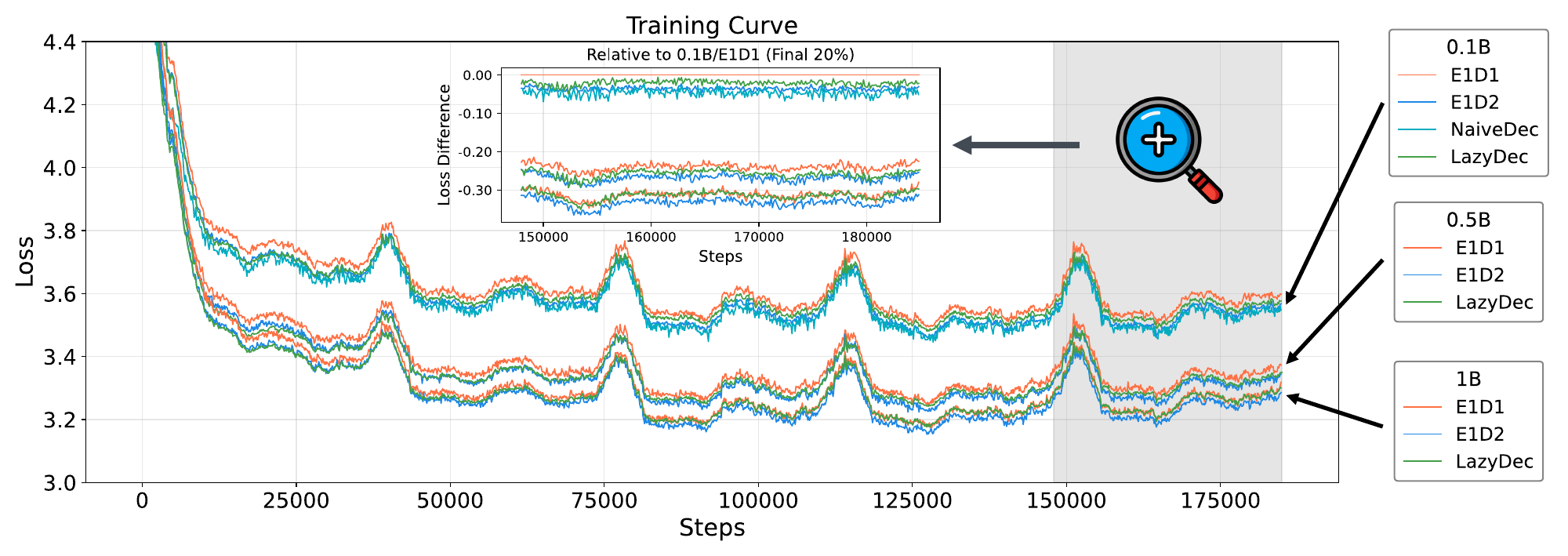}
    \caption{Training curves for different architectures across three model scales. Despite achieving similar loss, Lazy Decoder-Only architecture requires 10× fewer FLOPs than classic architectures. E1D1 and E1D2 denote encoder-decoder parameter ratios of 1:1 and 1:2, respectively.}
 \label{fig:arch_loss_curves}
\end{figure}

We compare three architectural paradigms for generative recommendation: the encoder-decoder architecture (OneRec-V1), the naive decoder-only architecture, and our proposed lazy decoder-only architecture. 
For each model, we evaluate the average generation loss across three semantic tokens:

\begin{equation}
\mathcal{L}_{\text{Gen}} = -\frac{1}{3} \sum_{i=1}^{3} \log p(s^i | \text{BOS}, s^{<i}, \text{Context}),
\end{equation}
where $s^i$ denotes the $i$-th semantic ID of the target item, $\text{BOS}$ represents the begin-of-sentence token, and $\text{context}$ is the output from the context processor including both user static and behavioral features.
This loss differs from OneRec-V1 as we use the average over three tokens, while V1 uses their sum.

Table~\ref{tab:arch_comparison} and Figure~\ref{fig:arch_loss_curves} present the computational requirements and convergence performance across different model scales. 
Despite requiring significantly fewer FLOPs and lower activation memory, our lazy decoder-only architecture achieves comparable losses compared to traditional approaches.

\subsubsection{Key-Value Sharing}
Our context processor introduces two key parameters that enable flexible control over the overall context dimensions: $L_{\text{kv}}$ and $S_{\text{kv}}$. 
The parameter $L_{\text{kv}}$ determines the number of distinct context representations across layers, with every $N_{\text{layer}} / L_{\text{kv}}$ consecutive decoder blocks sharing the same key-value pairs. 
The parameter $S_{\text{kv}}$ further controls whether keys and values share the same representation ($S_{\text{kv}} = 1$) or maintain separate projections ($S_{\text{kv}} = 2$).
This design reduces both computational cost and activation memory while maintaining comparable performance on the generative task. 
We conduct ablation studies on a 1B parameter dense lazy decoder model with $N_{\text{layer}} = 18$ to investigate the impact of these design choices.

\begin{table}[h]
\centering
\caption{Impact of key-value sharing strategies on model efficiency and performance. The number of activations is calculated under the batch size of 512.}
\label{tab:kv_sharing}
\begin{tabular}{ccccc}
\toprule
$\mathbf{L_{\text{kv}}}$ & $\mathbf{S_{\text{kv}}}$  & \textbf{GFLOPs} & \textbf{Activations} & \textbf{Convergence Loss} \\
\midrule
1 & 1 & 18.89 & 1.24B & 3.27 \\
1 & 2 & 19.19 & 1.33B & 3.27 \\
3 & 1 & 19.49 & 1.42B & 3.27 \\
9 & 1 & 21.27 & 1.99B & 3.27 \\
18 & 1 & 23.95 & 2.83B & 3.27 \\
\bottomrule
\end{tabular}
\end{table}

Figure~\ref{fig:kv_sharing_loss} demonstrates that aggressive key-value sharing maintains competitive loss throughout training, validating our efficient context processing strategy.

\subsubsection{Grouped Query Attention}

\begin{table}[h]
\centering
\caption{Impact of grouped query attention on model efficiency and performance. The number of activations and key-value size in cross-attention are calculated under the batch size of 512.}
\label{tab:gqa}
\begin{tabular}{ccccc}
\toprule
$\mathbf{G_{\text{kv}}}$ & \textbf{GFLOPs} & \textbf{Activations} & \textbf{KV Size} & \textbf{Convergence Loss} \\
\midrule
14 & 18.89 & 1.24B & 94M & 3.27 \\
7 & 18.74 & 1.19B & 47M & 3.28 \\
2 & 18.64 & 1.16B & 13M & 3.28 \\
1 & 18.62 & 1.15B & 7M & 3.27 \\
\bottomrule
\end{tabular}
\end{table}

Grouped Query Attention (GQA) shares key-value heads across multiple query heads. 
In our lazy decoder architecture, this optimization reduces activation memory and \textbf{memory access bottleneck} in cross-attention operations, thereby enhancing training throughput with minimal impact on  model quality.
We investigate the impact of varying the number of key-value head groups $G_{\text{kv}} \in \{1, 2, 7\}$ on a 1B parameter dense lazy decoder model with $14$ attention heads.

The results in Table~\ref{tab:gqa} and Figure~\ref{fig:gqa_loss} demonstrate that GQA with different number of groups yields nearly identical performance to full attention while substantially reducing memory requirements.

\subsubsection{Model Scaling}
\label{sec:scaling}
We conduct comprehensive scaling experiments on our lazy decoder-only architecture, investigating both dense and sparse configurations to understand the compute-performance trade-offs across different model scales.

\paragraph{Dense Model Scaling.} 
We explore the scaling properties of dense lazy decoder models ranging from 0.1B to 8B parameters. Table~\ref{tab:dense_scaling} presents the architectural hyperparameters and convergence performance for each model configuration.

\begin{table}[h]
\centering
\caption{Hyperparameter configurations and convergence loss for model scaling experiments.}
\renewcommand{\arraystretch}{1.2}
\resizebox{\textwidth}{!}{%
\begin{tabular}{c|c|c|c|c|c|c|c}
\toprule
\textbf{Model} & \textbf{Parameters} & \textbf{d\_model} & \textbf{n\_layers} & \textbf{n\_heads} & \textbf{embed\_dim} & \textbf{Learning Rate} & \textbf{Convergence Loss} \\
\midrule
\multirow{7}{*}{Dense} 
& 0.1B & 640 & 12 & 10 & 32 & 5.00e-4 & 3.57 \\
& 0.2B & 896 & 12 & 14 & 45 & 3.54e-4 & 3.46 \\
& 0.5B & 1408 & 14 & 11 & 70 & 2.24e-4 & 3.33 \\
& 1B & 1792 & 18 & 14 & 90 & 1.58e-4 & 3.27 \\
& 2B & 2304 & 22 & 18 & 115 & 1.12e-4 & 3.23 \\
& 4B & 2944 & 26 & 23 & 147 & 7.91e-5 & 3.20 \\
& 8B & 3584 & 34 & 28 & 179 & 5.59e-5 & 3.19 \\
\midrule
MoE & 4B (0.5B active) & 1408 & 14 & 11 & 70 & 2.24e-4 & 3.22 \\
\bottomrule
\end{tabular}%
}
\label{tab:dense_scaling}
\end{table}

\begin{figure}
    \centering
    \includegraphics[width=0.65\linewidth]{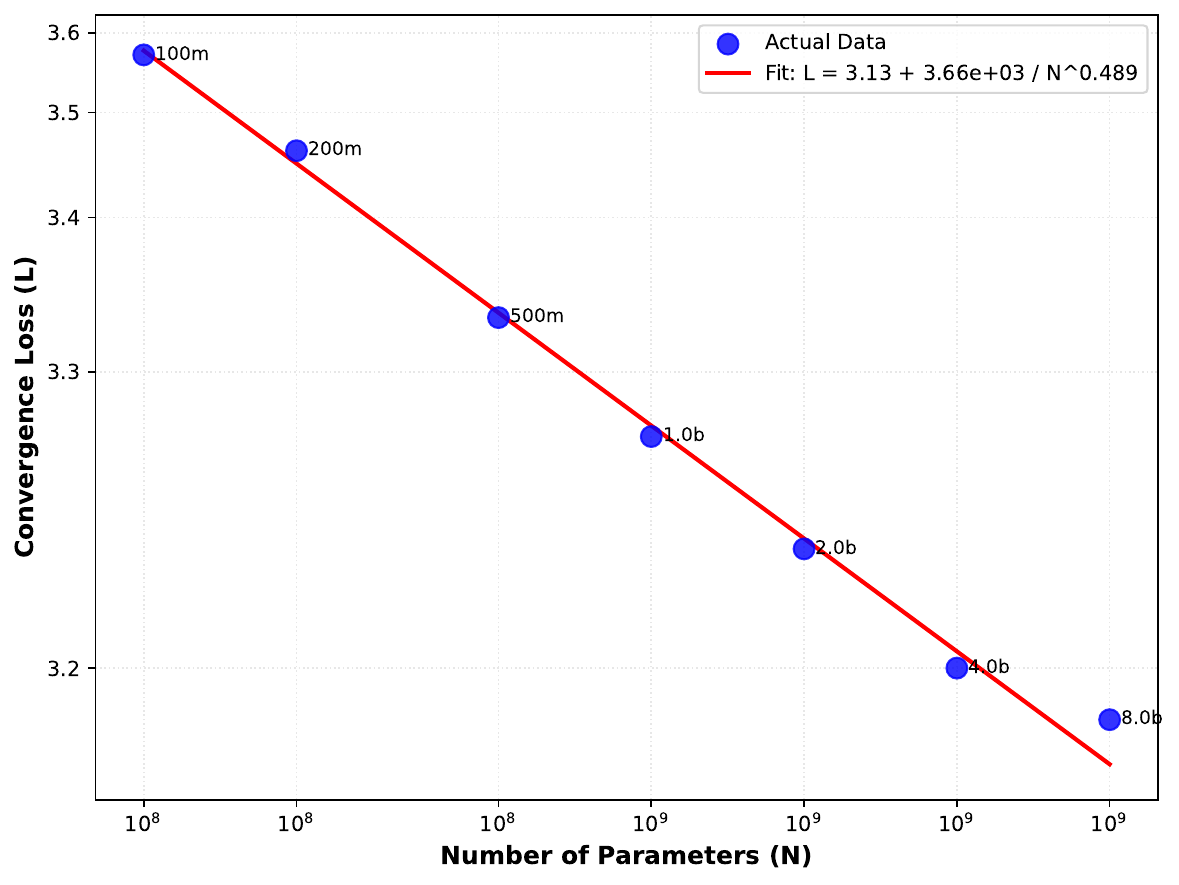}
    \caption{Dense Model Scaling Law Curve. The scaling law is fitted by $
        \hat{L}(N) \triangleq E + \frac{A}{N^\alpha}$, 
        where $E=3.13$, $A=3660$, $\alpha=0.489$.}
    \label{fig:dense_params_vs_loss}
\end{figure}

\paragraph{Sparse Mixture-of-Experts.} To achieve more efficient scaling, we investigate a Mixture-of-Experts (MoE) variant that replaces dense feed-forward networks with sparse expert routing. Our MoE configuration employs 53 routed experts and 1 shared expert, with total parameters of 4B (0.5B active per token). The model uses top-3 expert routing per token with an MoE intermediate size of 1408. 
The sparse model maintains the same base architecture as the 0.5B dense model while replacing feed-forward layers after the first 2 lazy decoder blocks with MoE layers.

\begin{figure}[t]
    \centering
    \includegraphics[width=1.\textwidth]{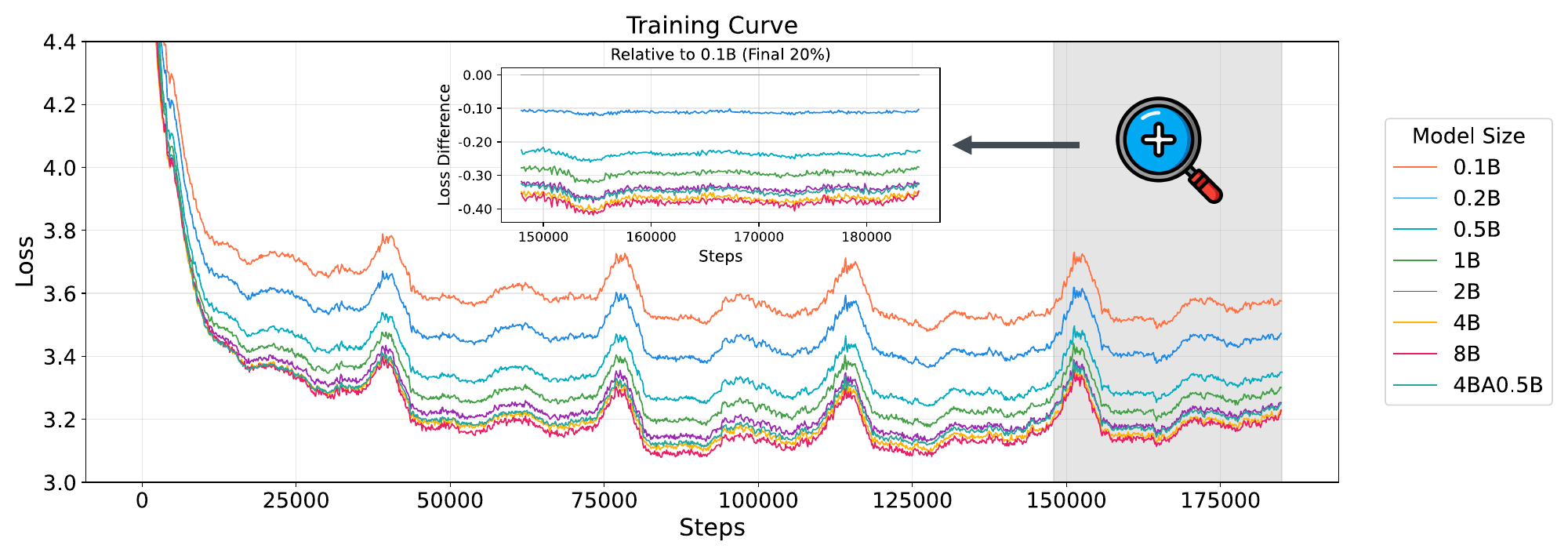}
    \caption{Training dynamics of lazy decoder architectures across model scales. Convergence loss decreases from 3.57 (0.1B) to 3.19 (8B). The 4B MoE variant (0.5B activated), denoted as 4BA0.5B in the figure, achieves competitive performance while maintaining computational efficiency.}
    \label{fig:scaling_comparison}
\end{figure}

\paragraph{Results and Analysis.} Figure~\ref{fig:scaling_comparison} illustrates the training dynamics across different model configurations. Our experiments reveal several key insights regarding the scaling behavior of lazy decoder architectures in recommendation systems. 
We also present how loss decreases as training budget increases for models with different scales, which can be found in the Figure~\ref{fig:training_budget}.

Our empirical results show reasonable agreement with the theoretical scaling law. 
While the general Chinchilla scaling law expression is
\[
\hat{L}(N, D) \triangleq E + \frac{A}{N^\alpha} + \frac{B}{D^\beta}
\]
where $N$ is the number of model parameters and $D$ is the number of training tokens~\cite{hoffmann2022training}, 
our experimental setup keeps $D$ fixed. In this case, the term $\frac{B}{D^\beta}$ is a constant and can be absorbed into the intercept $E$. 
Therefore, for fixed data, the scaling law simplifies to
\[
\hat{L}(N) \triangleq E' + \frac{A}{N^\alpha}
\]
where $E' = E + \frac{B}{D^\beta}$. Our empirical results fit this functional form very well:
$E=3.13$, $A=3660$, $\alpha=0.489$, as shown in Figure~\ref{fig:dense_params_vs_loss}.

The MoE variant with 4B total parameters (activating 0.5B) achieves a convergence loss of 3.22, outperforming the 2B dense model while maintaining computational requirements comparable to the 0.5B dense baseline. This configuration achieves a loss reduction of 0.11 compared to the 0.5B dense model, demonstrating the effectiveness of sparse architectures for recommendation tasks.

These results demonstrate that our lazy decoder architecture scales effectively, with MoE variants offering particularly attractive trade-offs for deployment in industrial-scale recommendation systems where computational efficiency directly impacts serving costs and latency.


\section{Preference Alignment with Real-World User Interactions
}
\label{Post-training}
In this section, we introduce the post-training phase of OneRec-V2.  
The Supervised Fine-Tuning phase is the same as in OneRec-V1, using streaming exposure data to perform online $\mathcal{L}_{Gen}$ loss training, consistent with the loss used during pretraining. The main purpose is to capture users' real-time interest changes while preventing the model from deviating too far from the pretrained model. In OneRec-V1, the RL phase was solely based on the reward model. In OneRec-V2, we introduce RL based on user feedback signals as rewards. 
\subsection{Reinforcement Learning with User Feedback Signals}
Defining rewards based on user feedback can avoid the issue of reward hacking and does not require additional model computation overhead.  
However, it still faces challenges such as how to combine multiple objectives and the sparsity of positive labels. In the short-video recommendation scenario, the playing time for each video is the densest feedback signal  
and is closely correlated with the most important online metrics, such as APP Stay Time and LT7 (Lifetime over 7 days).  
Therefore, we design a simple but effective reward based on playing time.

\subsubsection{Duration-Aware Reward Shaping}

\begin{figure}[ht]
    \centering
    \includegraphics[width=1.0\textwidth]{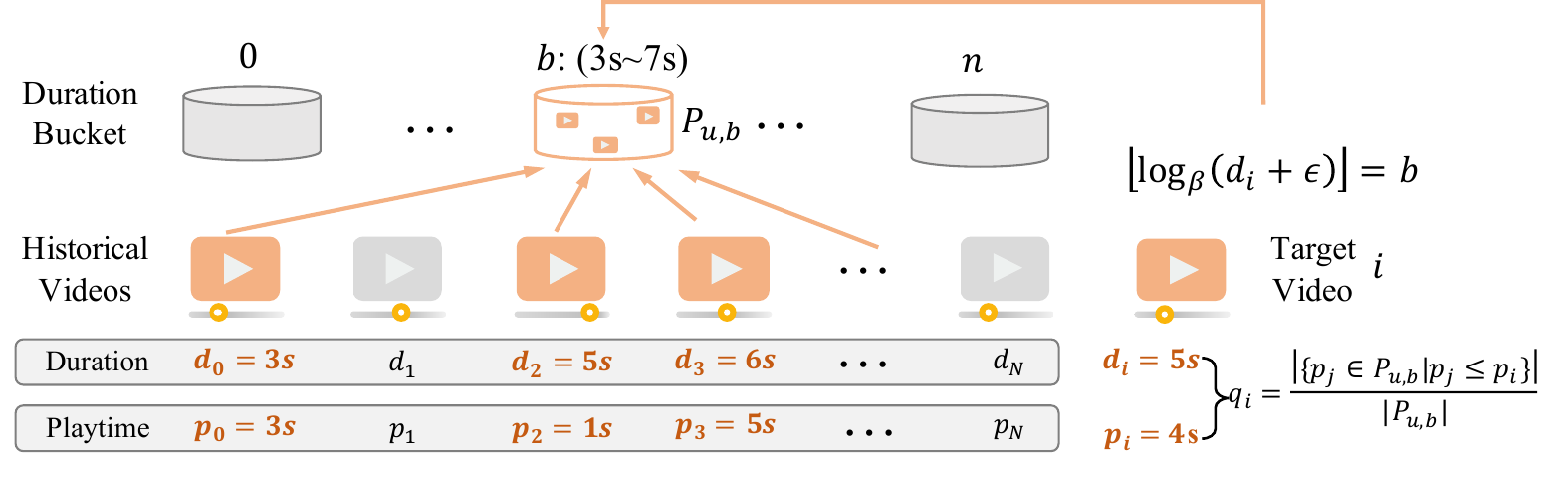}
    \caption{Illustration of the Duration-Aware Reward Shaping. The videos in a user's watch history are bucketed according to the durations, and for a target video, the quantile of its playing time within the corresponding bucket is computed as the user's preference score.
}
    \label{fig:reward}
\end{figure}
While video playing time is a useful indicator of user satisfaction, it is inherently biased by the duration of the video. To address this bias, we propose a \textit{Duration-Aware Reward Shaping} mechanism, as illustrated in Figure \ref{fig:reward}. The method normalizes playing time by comparing it with each user’s historical videos of comparable duration. Because video duration follows a long-tailed distribution, we partition historical videos into buckets using a logarithmic strategy. This approach groups durations into exponentially widening intervals, yielding more balanced and meaningful peer groups. The mapping is given by the function $\mathcal{F}(d)$, which assigns a video with duration $d$ to a discrete bucket index $b \in B$. Formally, the bucketing function is defined as:
\[
\mathcal{F}(d) = \lfloor \log_{\beta}(d + \epsilon) \rfloor
\]
where $\beta$ is a configurable logarithmic base controlling bucket granularity, and $\epsilon$ is a small constant (e.g., $10^{-6}$) added for numerical stability when processing very short durations.


Let $H_u = \{(d_k, p_k)\}_{k=1}^N$ denote the historical interaction sequence of user $u$, where $d_k$ is the video duration and $p_k$ the observed playing time. For each duration bucket $b$, we define the empirical distribution of playing times as
\[
P_{u,b} = \{p_j \mid (d_j, p_j) \in H_u, \mathcal{F}(d_j) = b\}.
\]
Given a target video $i$ with duration $d_i$ and playing time $p_i$, we first identify its bucket $b = \mathcal{F}(d_i)$. The duration-normalized engagement score is then computed as the empirical percentile rank of $p_i$ within the user’s historical distribution $P_{u,b}$:
\[
q_i = \frac{|\{p_j \in P_{u, b} \mid p_j \le p_i\}|}{|P_{u, b}|}.
\]

We select the most valuable samples as positive samples based on this score. In a batch, we compute $\tau_{b}$ as the 25\% quantile (top quartile) of $q_i$ after sorting them in descending order. For samples with explicit negative feedback such as "dislike" action ($neg_i = 1$), we set $A_i = -1$. All other samples are filtered out, which is equivalent to setting $A_i = 0$.  Note that we directly assign the advantage values without normalization,  because our definition of positive and negative samples is sufficiently strict. Further normalization may introduce inconsistency in optimization and thus degrade performance. Formally, the definition is as follows:
\[
A_i =
\begin{cases}
+1, & q_i > \tau_B \text{  and  } neg_i = 0, \\
-1, & neg_i = 1, \\
0,  & \text{otherwise}.
\end{cases}
\]
This strategy effectively filters for high-quality positive examples while incorporating direct negative signals, yielding a more accurate user preference signals.

\subsubsection{Reinforcement Learning}
\label{sec:rl}
\paragraph{Gradient-Bounded Policy Optimization} The effectiveness and stability of reinforcement learning have recently been a major research focus in the LLM community. A key challenge is to enhance exploration to improve performance while maintaining gradient stability. In this section, we introduce our newly proposed reinforcement learning method, \textbf{GBPO} (Gradient-Bounded Policy Optimization). 

\begin{equation}
\label{eq:GBPO}
\mathcal{J}_{GBPO}(\theta) = -\mathbb{E}_{u\sim P(U),\{o_i\}_{i=1}^G\sim\pi_{\theta_{old}}}\left[\frac1G\sum_{i=1}^G\frac{\pi_{\theta}(o_i|u)}{\pi_{\theta_{old}}'(o_i|u)}\cdot A_i\right],    
\end{equation}

\begin{equation}
    \pi_{\theta_{old}}'(o_i|u)=
    \begin{cases}
       \text{max}(\pi_{\theta_{old}}, sg(\pi_{\theta})),&A_i\geq0, \\[6pt]
       \text{max}(\pi_{\theta_{old}}, 1-sg(\pi_{\theta})),&A_i< 0. \\
       
    \end{cases}
\end{equation}
From the formulation, we can see that GBPO removes the clipping operation on the ratio and introduces a dynamic bound on $\pi_{\theta_{old}}$.  
Overall, GBPO has two main strengths:
\begin{itemize}
    \item \textbf{Full Sample Utilization}: preserving the gradients of all samples, encouraging the model to perform more diverse exploration.  
    \item \textbf{Bounded Gradient Stabilization}: bounding the gradients of RL with the gradients of the BCE (Binary Cross-Entropy) loss, enhancing the stability of RL training. 
\end{itemize}

\paragraph{Existing Clipping-based Work} Before detailing \textbf{GBPO}, we first briefly review existing RL methods for LLMs. GRPO/PPO~\citep{shao2024deepseekmathgrpo, schulman2017proximalppo} discard samples with excessively large or small policy ratios through a clipping operation, preventing overly aggressive training. DAPO~\citep{yu2025dapo} relaxes sample restrictions via clip higher, especially by incorporating more low-probability or high-entropy tokens, thereby increasing diversity while improving reinforcement learning performance. These studies indicate that relaxing clipping constraints to include more samples can encourage more diverse exploration and improve performance. 

However, these methods do not provide a complete and comprehensive consideration of gradient stability. In particular, for negative samples, the absence of an upper bound on the policy ratio can easily lead to gradient explosion, causing the model’s performance to collapse. Dual-clip ~\citep{ye2020mastering} applies an upper bound truncation to the policy ratio for negative samples. While this improves stability, it discards too many negative samples, which slows the convergence. ECPO~\citep{zhou2025onerec} mitigates gradient explosion on negative samples by applying clipping directly to $\pi_{old}$ rather than to the ratio $\pi_\theta / \pi_{old}$. This strategy retains a larger proportion of training samples while improving optimization stability. Similarly, CISPO~\citep{chen2025minimax} and GPPO~\citep{su2025klear} adopt related techniques to keep the ratio within a reasonable range, while preserving gradient signals from more samples. In OneRec V1, we employ ECPO (Early Clipped GRPO), formally defined as:

\begin{equation}
\label{eq:ecpo}
\mathcal{J}_{ECPO}(\theta) = -\mathbb{E}_{u\sim P(U),\{o_i\}_{i=1}^G\sim\pi_{\theta_{old}}}\left[\frac1G\sum_{i=1}^G\text{min}\left(\frac{\pi_{\theta}(o_i|u)}{\pi_{\theta_{old}}'(o_i|u)}A_i,\text{clip}\left(\frac{\pi_{\theta}(o_i|u)}{\pi_{\theta_{old}}'(o_i|u)},1-\epsilon,1+\epsilon\right)A_i\right)\right],    
\end{equation}

\begin{equation}
    \pi_{\theta_{old}}'(o_i|u)=\text{max}\left(\frac{\text{sg}(\pi_\theta(o_i|u))}{1+\epsilon+\delta},\pi_{\theta_{old}}(o_i|u)\right), \quad\delta>0.
\end{equation}

\begin{figure}[h!]
\centering
\includegraphics[width=.42\textwidth]{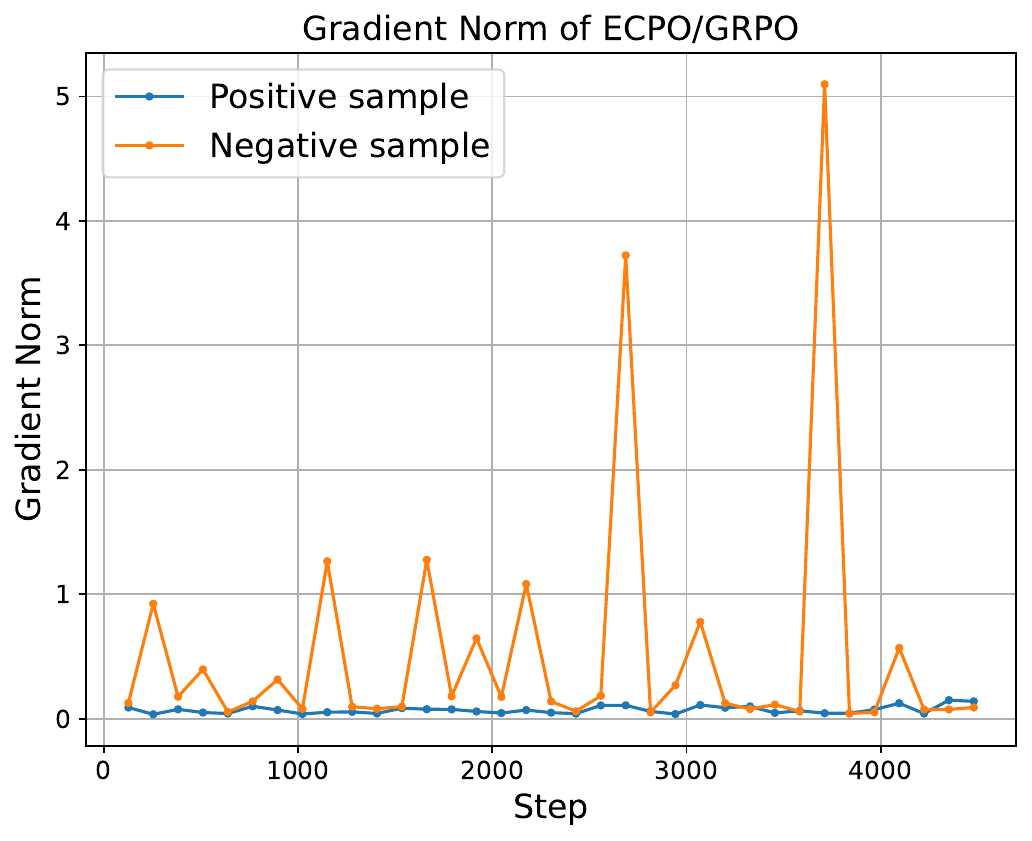}
\includegraphics[width=.437\textwidth]{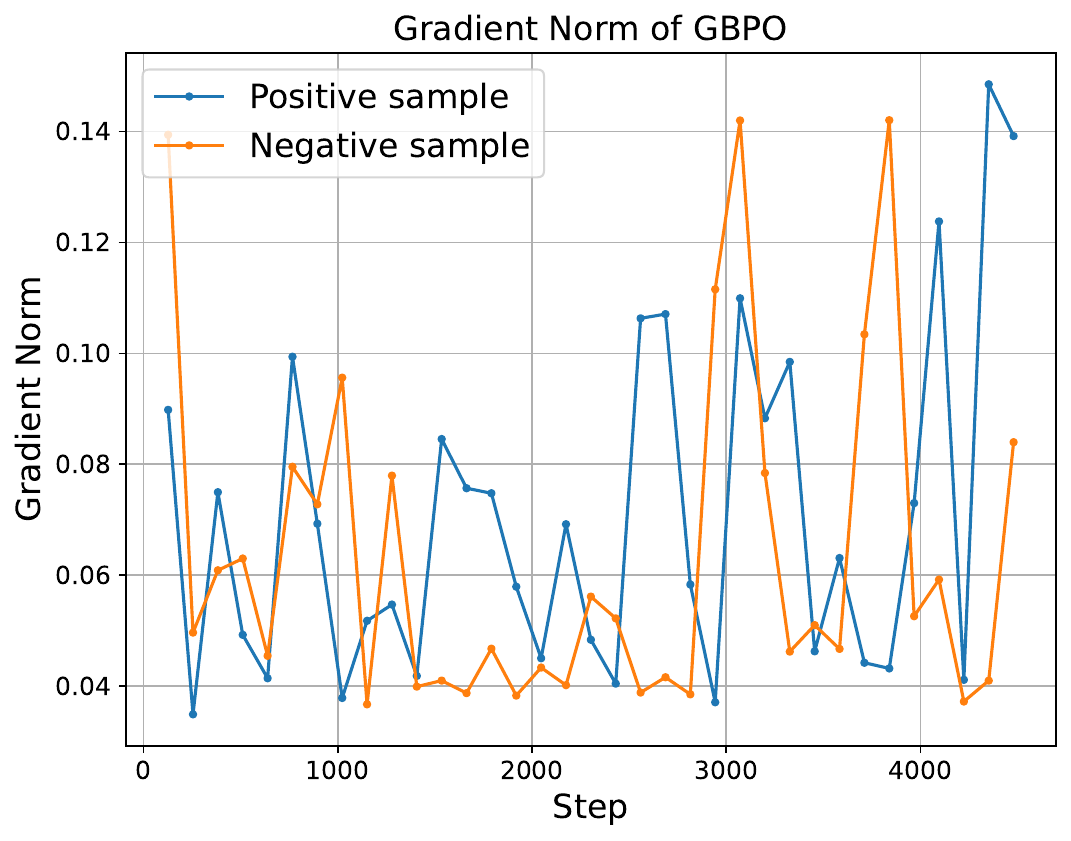}
\caption{Gradient comparison between GBPO and traditional ratio-clipping methods. In training of negative samples, GBPO exhibits significantly more stable gradients.}
\label{fig:grad}
\end{figure}

\paragraph{Gradient Analysis} The exposure samples include both those generated by OneRec and those from the traditional pipeline. For exposure samples generated by OneRec, we use the generation probability at the time of exposure as $\pi_{old}$. For samples from the traditional pipeline, due to the complexity of the pipeline, we cannot obtain their probabilities; therefore, we simplify $\pi_{old}$ to the current generation probability of the OneRec model, i.e., $\pi_{old} = \mathrm{sg}(\pi_{\theta})$. For these samples, the policy ratio is always $1$. In traditional RL methods, samples with a ratio of $1$ are regarded as stable for training and are not subjected to truncation. However, in reality, such samples can still cause gradient explosion, which is induced by negative samples, as shown in Figure~\ref{fig:grad}.

\begin{figure}[ht]
\centering
\includegraphics[width=1.0\textwidth]{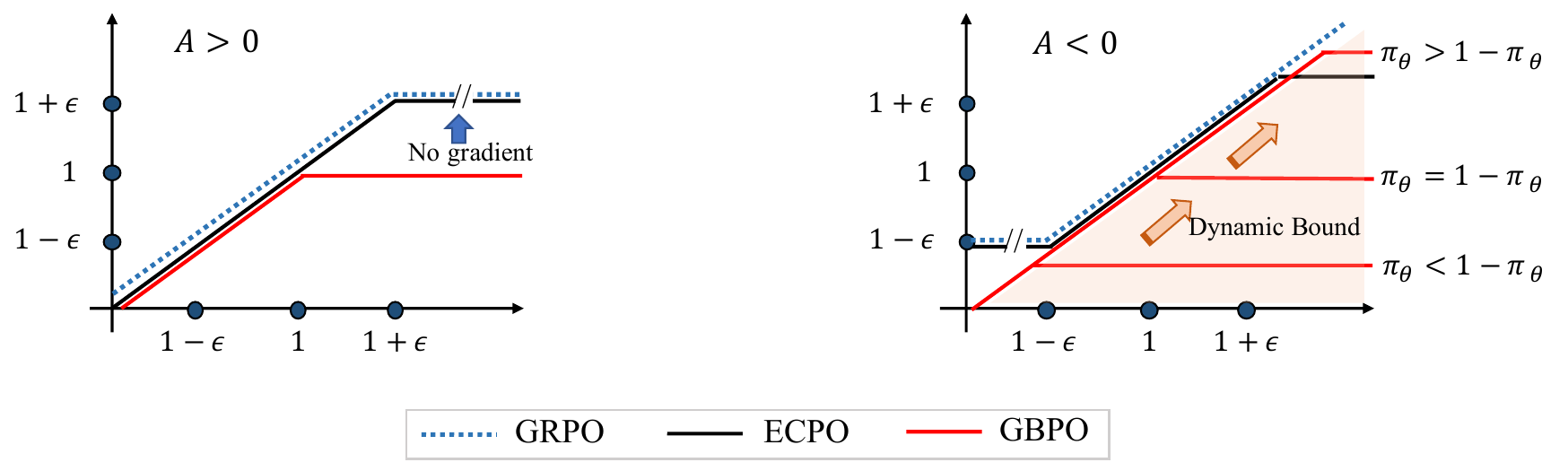}
\caption{Illustration of GBPO. The $x$-axis is $\pi_{\theta} / \pi_{\theta_{old}}$ and the $y$-axis is the clipped $\pi_{\theta} / \pi_{\theta_{old}}$. "//" means "No gradient". Compared with traditional ratio-clipping methods, the main differences of GBPO are: 1. It does not discard the gradients of any samples. 2. For negative samples, the bounding of the ratio is based on a dynamic bound related to $\pi_{\theta}$.}
\label{fig:gbpo}
\end{figure}

From the perspective of gradients, for a specific token $i$ of these samples, we have  
\begin{equation}
     \mathcal{J}_{ECPO}^i(\theta)= -A_i \cdot \frac{\pi_{\theta}}{sg(\pi_\theta)},
\end{equation}
\begin{equation}
 \frac{\partial \mathcal{J}_{ECPO}^i(\theta)}{\partial \theta} = - A_i\cdot\frac{1}{\pi_\theta} \frac{\partial \pi_\theta}{\partial \theta},   
\end{equation}
which indicates that the smaller the current token probability $\pi_\theta$, the larger the gradient.  
For positive samples, a smaller probability means there is more room to boost it, so having a larger gradient is reasonable.  
However, for negative samples, a smaller probability means there is less room to suppress it; if the gradient is too large, it can easily lead to model overfitting or even collapse. This phenomenon indicates that traditional clipping methods cannot fully resolve the issue of unstable RL gradients, as they cannot avoid gradient explosion when the ratio is $1$. In the BCE loss, there is likewise a penalty for negative samples, but its gradients are much more stable compared to those of the RL loss.

\begin{equation}
    \mathcal{L}_{BCE}(y, p_\theta) = - \left[ y \cdot \log(p_\theta) + (1-y) \cdot \log(1-p_\theta) \right],
\end{equation}
\begin{equation}
    \frac{\partial \mathcal{L}_{BCE}}{\partial \theta}=
    \begin{cases}
    -\dfrac{1}{p_\theta}\dfrac{\partial p_\theta}{\partial \theta}, & y=1,\\[6pt]
    \dfrac{1}{1-p_\theta}\dfrac{\partial p_\theta}{\partial \theta}, & y=0.
    \end{cases}
\end{equation}
For negative samples, the smaller the current model probability, the smaller the gradient when suppressing them,  
leading to a more stable model.  
Based on this observation, we propose \textbf{GBPO}, which bounds the RL gradients with the more stable gradients from the BCE loss. We illustrate the differences in Figure \ref{fig:gbpo}.

\subsubsection{Experiment}
\label{sec:exp_post_training}
\paragraph{Experiment Settings}
\label{sec:3.1.3}
In this section, we experimentally verify the effectiveness of the defined user feedback signals. For rapid validation, all experiments in this section are conducted under the setting of the 0.5b model with a context length of 512. The baseline is OneRec-V1. In the experimental setting of OneRec-V1, the online traffic allocated to the experimental group was only a very small fraction of the total, so the training samples were drawn almost entirely from the traditional recommendation pipeline. In the LLM domain, it has been shown that training on self-generated samples can lead to self-improvement \citep{he2025skywork}. With OneRec now serving 25\% of total traffic, we have sufficient data to validate this hypothesis in our setting. Accordingly, we design two experimental groups for comparison:

\vspace{-0.6em}
\begin{itemize}[leftmargin=*]
\item \textbf{w/o OneRec Samples}: Using only samples generated by the traditional recommendation pipeline for reinforcement learning, which aligns the samples with OneRec-V1.

\item \textbf{w/ OneRec Samples}: Incorporating samples generated by the OneRec pipeline, which also include the samples generated by the current model's experimental group. In other words, this setting introduces on-policy reinforcement learning.
\end{itemize}
As mentioned before, positive samples for reinforcement learning are identified as the top 25\% percentile of videos ranked by the duration-aware reward score, while negative samples are identified by explicit negative feedback (e.g., a "dislike" action). Note that the total number of training samples is kept essentially the same across the two groups. The reinforcement learning loss is GBPO (Equation~\ref{eq:GBPO}). All the results are presented in Table~\ref{tab:post_training_results}.

\paragraph{Result Analysis} From Table \ref{tab:post_training_results}, we have the following observations. When using only samples from the traditional pipeline, i.e., with the same sample source as OneRec-V1, introducing user-feedback-based reinforcement learning significantly improves duration-related metrics such as App Stay Time and Watch Time, but replaces some other metrics such as Video View. This indicates that our duration-aware reward is indeed strongly correlated with App Stay Time. After incorporating samples from the OneRec pipeline, almost all metrics improve significantly, with Video View in particular turning from negative to positive. This demonstrates that user-feedback-based reinforcement learning enables self-iterative optimization, fully leveraging user feedback signals to enhance user experience.
\begin{table*}[h]
    \centering
    \caption{Online A/B testing results for user feedback signals based RL. All metrics show relative improvements over the OneRec-V1 baseline.}
    \renewcommand{\arraystretch}{1.2}
    \setlength{\tabcolsep}{8pt}
    \begin{tabular}{p{2.5cm}|l|cc}
    \toprule
    \textbf{Scenarios} & \textbf{Online Metrics} & \textbf{w/o OneRec Samples} & \textbf{w/ OneRec Samples} \\
    \hline
    \multirow{8}{*}{Kuaishou} & App Stay Time & +0.165\% & +0.227\% \\
    & Watch Time & +1.054\% & +0.648\% \\
    & Video View & -0.901\% & +0.716\% \\
    & Like & -0.186\% & +2.897\% \\
    & Follow & +2.274\% & +3.661\% \\
    & Comment & -4.982\% & +6.392\% \\
    & Collect & -0.817\% & +1.232\% \\
    & Forward & -2.162\% & +3.426\% \\
    \hline
    \multirow{8}{*}{Kuaishou Lite} & App Stay Time & +0.159\% & +0.353\% \\
    & Watch Time & +0.396\% & +0.104\% \\
    & Video View & -2.231\% & +0.575\% \\
    & Like & -0.534\% & +4.956\% \\
    & Follow & +1.809\% & +4.800\% \\
    & Comment & -4.860\% & +5.067\% \\
    & Collect & -0.377\% & +2.701\% \\
    & Forward & +0.775\% & +5.783\% \\
    \bottomrule
    \end{tabular}
    \label{tab:post_training_results}
\end{table*}

\subsection{User Feedback Signals versus Reward Model}
\subsubsection{Limitation of Reward Model}
In this section, we compare reinforcement learning in OneRec-V1, which relies on a reward model, with reinforcement learning driven by user feedback signals. Although OneRec-V1 demonstrated the effectiveness of reinforcement learning through extensive experiments, its performance was constrained by limited sampling probability. Due to resource constraints, on-policy roll-outs could only be conducted for a small subset of users (1\%). Moreover, the reward model is susceptible to \emph{reward hacking}. User feedback signals directly reflect real user preferences, thereby mitigating the risk of reward hacking. However, before the full deployment of OneRec, large-scale real user feedback on generated samples was not available. With the full deployment of OneRec, these signals can now be leveraged more effectively for precise self-iterative optimization. In the previous section, we demonstrated the effectiveness of the proposed duration-aware feedback signals. Now, we compare the performance of user feedback with that of the reward model.

\begin{table*}[h]
    \centering
    \caption{Online A/B testing results for RL training of OneRec-V2.}
    \renewcommand{\arraystretch}{1.2}
    \setlength{\tabcolsep}{7pt}
    \begin{tabular}{p{2.5cm}|l|ccc}
    \toprule
    \textbf{Scenarios} & \textbf{Online Metrics} & \textbf{Reward Model} & \textbf{User Feedback Signals} & \textbf{Hybrid} \\
    \hline
    \multirow{8}{*}{Kuaishou} & App Stay Time & +0.269\% & \textbf{+0.299\%} & +0.283\% \\
    & Watch Time & +0.537\% & \textbf{+0.610\%} & +0.118\% \\
    & Video View & +0.505\% &\textbf{+0.647\%} & \textbf{+0.647\%} \\
    & Like & +6.552\% & +2.435\% & \textbf{+7.010\%} \\
    & Follow & +7.265\% & +2.007\% & \textbf{+8.458\%} \\
    & Comment & \textbf{+15.472\%} & +0.944\% & +8.763\% \\
    & Collect & +1.856\%& +1.401\% &  \textbf{+9.739\%}  \\
    & Forward & \textbf{+12.024\%} &+0.803\%  &  +5.270\%\\
    \hline
    \multirow{8}{*}{Kuaishou Lite} & App Stay Time & +0.163\% & \textbf{+0.213\%} & +0.207\% \\
    & Watch Time & \textbf{+0.503\%} & +0.172\% & -0.398\% \\
    & Video View & \textbf{+0.457\%} & +0.056\% & +0.083\% \\
    & Like & \textbf{+7.798\%} & +4.008\% & +6.267\% \\
    & Follow & +12.242\% & +4.421\% & \textbf{+11.705\%} \\
    & Comment & \textbf{+11.284\%} & +3.958\% & +7.002\% \\
    & Collect & \textbf{+4.468\%} & +1.731\%  & +3.495\% \\
    & Forward & \textbf{+15.919\%} & +7.704\% & +6.670\%\\
    \bottomrule
    \end{tabular}
    \label{tab:rm_exp}
\end{table*}

\subsubsection{Experiment}
\paragraph{Experiment Settings} We set up three groups of experiments for comparison,  
referred to as \textbf{Reward Model}, \textbf{User Feedback Signals}, and \textbf{Hybrid}. The model setting is the same as in section \ref{sec:3.1.3}.
The evaluation metrics are the same as in the previous experiments,  
including both duration-based metrics and interaction-based metrics.  
The \emph{App stay time} is the most important metric,  
while the other metrics serve as reference values for user experience. The results shown in Table~\ref{tab:rm_exp} represent the relative performance of each group compared to OneRec-V1.

\vspace{-0.6em}
\begin{itemize}[leftmargin=*]
\item \textbf{Reward Model}: Introducing reward-model-based reinforcement learning,  
the main difference from OneRec-V1 lies in the architecture of the pretrained generative model. OneRec-V1 uses an Encoder-Decoder architecture, whereas OneRec-V2 employs the proposed Lazy Decoder.
\item \textbf{User Feedback Signals}: Introducing user-feedback-based reinforcement learning and incorporating self-generated samples, which is same as the "w/ OneRec Samples" setting in the last section.
\item \textbf{Hybrid}: Simultaneously introducing both reward model and user feedback signals, with the two types of samples being independent: the former are samples obtained through the model’s own rollout sampling, while the latter are samples that were previously exposed to users.

\end{itemize}


\begin{table*}[t]
    \centering
    \caption{The relative improvement of OneRec-V2 compared to OneRec-V1 in the online A/B testing.}
    \renewcommand{\arraystretch}{1.2} 
    \setlength{\tabcolsep}{20pt}      
    \begin{tabular}{p{2.5cm}|l|c}
    \toprule
    \textbf{Scenarios} & \textbf{Online Metrics} & \textbf{ OneRec-V2} \\
    \hline
    \multirow{9}{*}{Kuaishou} & App Stay Time & +0.467\% \\
    & LT7 & +0.069\% \\
    & Watch Time & +1.367\% \\
    & Video View & +0.331\% \\
    & Like & +3.924\% \\
    & Follow & +4.730\% \\
    & Comment & +5.394\% \\
    & Collect & +2.112\% \\
    & Forward & +3.183\% \\
    \hline
    \multirow{9}{*}{Kuaishou Lite} & App Stay Time & +0.741\% \\
    & LT7 & +0.034\% \\
    & Watch Time & +0.762\% \\
    & Video View & +0.259\% \\
    & Like & +5.393\% \\
    & Follow & +5.627\% \\
    & Comment & +5.013\% \\
    & Collect & +3.202\% \\
    & Forward & +7.958\% \\
    \bottomrule
    \end{tabular}
    \label{online_ab_table1}
\end{table*}

\paragraph{Results Analysis} From Table \ref{tab:rm_exp}, we can summarize the following observations. 
\vspace{-0.6em}
\begin{itemize}[leftmargin=*]
\item In the reward-model setting, OneRec-V2 performs significantly better than OneRec-V1, further confirming the advantages brought by the Lazy Decoder architecture. 

\item Whether based on the reward model or user feedback, reinforcement learning yields dual gains in both duration and interaction metrics. However, the reward model tends to favor improvements in interaction metrics, while real user feedback tends to favor increases in App Stay Time.  
This is because the rewards output by the reward model are a fusion of multiple recommendation objectives,  
whereas the rewards we define based on user feedback are computed mainly from video playing time. This also indicates that different reward definitions lead to different model preferences, which is consistent with the conclusions in OneRec-V1. 

\item When combining the two (Hybrid), although the specific gains in duration and interaction are not as high as those from each individual strategy, the loss in performance is minimal, and the balance between App Stay Time and interaction metrics is improved. This is because the gains brought by the two individual strategies partially overlap. Although combining them cannot achieve a perfect additive effect, it allows them to complement each other.  
This also highlights the importance of diversified reward signals. We will conduct further research on the diversity and accuracy of reward signals in the future.
\end{itemize}

\section{Online A/B Test}
\label{online_ab_test}
\label{sec:online}
We deployed OneRec-V2 across two major short-video scenarios on Kuaishou: the main Kuaishou feed and Kuaishou Lite feed, which represent the platform's highest-traffic environments serving 400 million daily active users. The evaluation was conducted using a 5\% traffic experimental group over a one-week observation period. The model used was a 1B-parameter version with a context length of 3000 and a beam size of 512. For online inference, the system utilized L20 GPUs and achieved a latency of 36ms and an MFU (Model FLOPs Utilization) of 62\%. To reduce system complexity, this version incorporated only User Feedback Signals. Our primary evaluation metrics were App Stay Time (measuring total user engagement duration) and LT7 (7-day user lifetime retention). As demonstrated in Table \ref{online_ab_table1}, OneRec-V2 achieved substantial improvements across both platforms. Furthermore, OneRec-V2 exhibited significant gains across all user interaction metrics, including likes, follows, comments, and other engagement behaviors, demonstrating its capability to guide multi-task recommendation systems toward a more balanced equilibrium while effectively mitigating seesaw effects between competing objectives. 

To further validate our findings, we conducted an additional experiment with caching disabled where all traffic in a separate 1\% experimental group requests OneRec-V2 (detailed results in Appendix~\ref{caching_disabled}). This comprehensive evaluation confirms the substantial improvements in user engagement metrics, with interaction indicators such as likes, follows, comments, and forwards showing remarkable gains of 9.6\% to 29.2\% across platforms. While these results demonstrate OneRec-V2's strong performance in driving user engagement, they also reveal important ecosystem-level considerations, including significant reductions in cold-start video views (44.7\% and 36.7\% for Kuaishou and Kuaishou Lite respectively) and increased cluster density. 
\section{Conclusion, Limitations, and Future Directions}
\label{Conclusion}
In this paper, we introduce OneRec-V2, building upon the foundation of OneRec-V1. We delve into the design of its scaling and reward systems. Regarding scaling, we found that although the OneRec-V1 model utilized MoE to allocate a large number of parameters in the decoder, the context encoding process consumed most computational resources due to sequence length differences, hindering further scalability and performance. Consequently, we rethought the model architecture and proposed a lazy decoder-only architecture, which shifts computation to the decoding phase, allowing for further model expansion (currently scaling to 8B). Additionally, we developed a method that effectively utilizes real user feedback to align user preferences. Unlike V1, which solely relied on a reward model for alignment, we incorporated real user feedback signals and, through innovative design, established a correlation between short-term video watching time and long-term satisfaction. Furthermore, using GBPO, we achieved highly stable training. Rigorous A/B experiments  have proven the effectiveness of this framework. However, our system still has room for improvement. For example:

\begin{enumerate}
\item Scaling: As the model scaled from 0.1B to 8B parameters, we observed a consistent decrease in loss, which is remarkably well described by the empirical scaling law proposed by Hoffmann et al. (2022)~\citep{hoffmann2022training}. Our results show excellent agreement with this scaling relation, as evidenced in Figure~\ref{fig:dense_params_vs_loss}. This validates the effectiveness of the chosen architecture and indicates that further scaling and architectural innovations can potentially yield even better performance.

 \item Reward System: We have newly incorporated real user feedback into the reward system, which has proven effective. However, our current solution establishes rules linking short-term and long-term returns, rather than allowing the model to directly optimize its long-term value. We will optimize in this direction to enable the model to achieve self-reinforcement towards long-term value.

\end{enumerate}

In addition to achieving profitability in video recommendation of Kuaishou  platform, OneRec-V2 has been deployed in various business scenarios, generating substantial returns, e.g.,~\citep{wei2025oneloc}. We believe this system can be further improved with iteration, verification, and optimization by more researchers and engineers.
\newpage

\bibliographystyle{abbrvnat}
\nobibliography*
\bibliography{bibtex}

\clearpage

\appendix
\section*{Appendix}
\section{Contributions}
Within each role, authors are listed alphabetically by their first name. 

\begin{multicols}{3}
\noindent
\textbf{ Core Contributors} \\
 Guorui Zhou \\
 Hengrui Hu \\
 Hongtao Cheng \\
 Huanjie Wang \\
 Jiaxin Deng \\
 Jinghao Zhang \\
 Kuo Cai \\
 Lejian Ren \\
 Lu Ren \\
 Liao Yu \\
 Pengfei Zheng \\
 Qiang Luo \\
 Qianqian Wang \\
 Qigen Hu \\
 Rongzhou Zhang \\
 Rui Huang \\
 Ruiming Tang \\
 Shiyao Wang \\
 Shujie Yang \\
 Tao Wu \\
 Wuchao Li \\
 Xinchen Luo \\
 Xingmei Wang \\
 Yi Su \\
 Yunfan Wu \\
 Zexuan Cheng \\
 Zhanyu Liu \\
 Zixing Zhang

\noindent
\textbf{ Contributors} \\
 Bin Zhang\\
 Boxuan Wang \\
 Chaoyi Ma \\
 Chengru Song \\
 Chenhui Wang \\
 Chenglong Chu \\
 Di Wang \\
 Dongxue Meng\\
 Dunju Zang \\
 Fan Yang \\
 Fangyu Zhang \\
 Feng Jiang \\
 Fuxing Zhang \\
 Gang Wang \\
 Guowang Zhang \\
 Han Li \\
 Honghui Bao \\
 Hongyang Cao \\
 Jiaming Huang \\
 Jiapeng Chen \\
 Jiaqiang Liu \\
 Jinghui Jia \\
 Kun Gai\\
 Lantao Hu \\
 Liang Zeng \\
 Qiang Wang \\
 Qidong Zhou\\
 Shengzhe Wang \\
 Shihui He \\
 Shuang Yang \\
 Siyang Mao \\
 Sui Huang \\
 Tiantian He \\
 Tingting Gao\\
 Wei Yuan \\
 Xiao Liang \\
 Xiaoxiao Xu \\
 Xugang Liu \\
 Yan Wang\\
 Yang Zhou \\
 Yi Wang \\
 Yiwu Liu \\
 Yue Song \\
 Yufei Zhang \\
 Yunfeng Zhao \\
 Zhixin Ling \\
 Ziming Li \\
\end{multicols}

\section{Computational Complexity of Different Architecture}
\label{appendix:computation_ana}
\paragraph{Preliminary.}
In practical recommender systems, multiple items are impressed simultaneously.
A key optimization is common context compression: when impressing $\mathbf{k}$ item recommendations to the same user, the shared contextual information (user profile, historical behaviors) needs to be processed only once and can be reused across all target items. 
This reduces the effective context length from $\mathbf{N}$ to approximately $\mathbf{N/k}$ tokens per item.
In KuaiShou, $k=5$.

The main computational components in a transformer block~\citep{vaswani2017attention} include: (1) feed-forward networks (FFNs), (2) attention projections ($W_q$, $W_k$, $W_v$, $W_o$), and (3) attention score computation. Their computational complexities are:
\begin{align}
    \text{FFN:} &\quad O(L \cdot d_{\text{model}} \cdot d_{\text{ff}}) \approx O(L \cdot 4d_{\text{model}}^2) \\
    \text{Attention Projections:} &\quad O(L \cdot 4d_{\text{model}}^2) \\
    \text{Attention Scores:} &\quad O(L^2 \cdot d_{\text{model}})
\end{align}
where $L$ is the number of tokens processed by these modules and $d_{\text{model}}$ is the model's hidden dimension. 
Notably, both FFN and attention projections can be approximated as $O(L \cdot D)$, where $D$ is the corresponding module's parameter count.


\paragraph{Encoder-Decoder Architecture.} 
We analyze the computational requirements of an encoder-decoder model with 0.5B parameters in both encoder and decoder components. 
During training with compressed context length $N/5$, the floating-point operations (FLOPs) decompose as follows:

\begin{align}
    \text{Context Transformation (Encoder):} &\quad 6 \times 0.5\text{B} \times \frac{N}{5} = 0.6N \text{ GFLOPs}\\
    \text{Context Projection (Cross-Attention):} &\quad 6 \times 0.05\text{B} \times \frac{N}{5} = 0.06N \text{ GFLOPs} \\
    \text{Context Decoding:} &\quad 0.6N + 0.06N = 0.66N \text{ GFLOPs} \\
    \text{Target Decoding:} &\quad 6 \times 0.45\text{B} \times 3 = 8.1 \text{ GFLOPs} \\
    \text{Total Computation:}  & \quad 0.66N + 8.1\text{ GFLOPs} \label{eq:total_comp}
\end{align}

\noindent where the factor of 6 accounts for both multiply-accumulate operations (contributing a factor of 2) and the forward-backward pass ratio (contributing a factor of 3). The context projection matrices ($W_k$, $W_v$) in the cross-attention mechanism reside within the decoder, comprising approximately 10\% (0.05B) of the decoder's parameters.

The computations of attention scores are ignored here.
Consider the specific model configuration with 9 encoder and 9 decoder layers, and $d_{\text{model}} = 1792$. 
The attention score computations are, Encoder: $6 \times  9 \times (\frac{N}{5})^2 \times 1792 = 3.8N^2$ KFLOPs, Decoder: $6 \times 9 \times 3 \times N \times 1792 = 290N$ KFLOPs.
When $N=512$, these values are orders of magnitude smaller than FFNs and attention projections.

\paragraph{Naive Decoder-Only Architecture.} 
For a decoder-only model with 1B parameters processing $N/5 + 3$ tokens with causal attention masking:
\begin{align}
    \text{Context Decoding:} &\quad  6 \times 1\text{B} \times \frac{N}{5} = 1.2N \text{ GFLOPs} \\
    \text{Target Decoding:} &\quad 6 \times 1\text{B} \times 3 = 18 \text{ GFLOPs} \\
    \text{Total Computation:} &\quad 1.2N + 18 \text{ GFLOPs} 
\end{align}

\section{Empirical Results}
\begin{figure}[h]
    \centering
    \includegraphics[width=1.\textwidth]{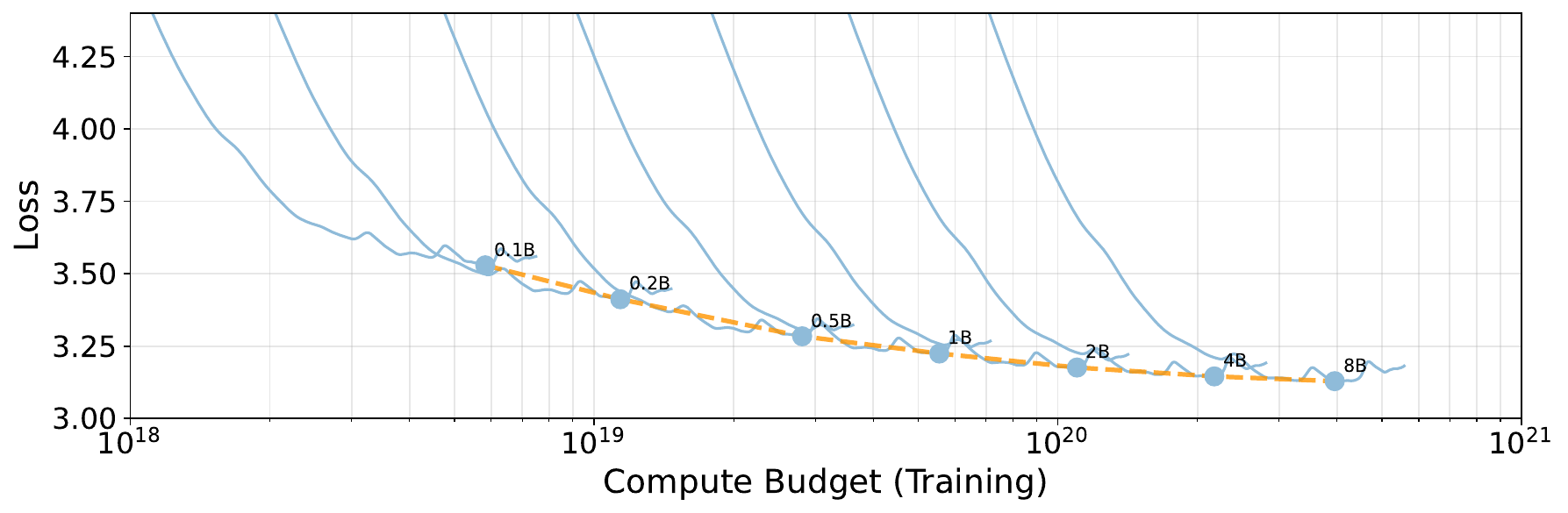}
    \caption{Smoothed generative training loss curves for OneRec-V2 models as a function of total compute (measured in FLOPs), demonstrating the scaling behavior and convergence patterns across model sizes. The orange line connects the minimum loss points achieved by different models.}
    \label{fig:training_budget}
\end{figure}

We conduct experiments to investigate the relationship between model size, compute budget, and the training loss for OneRec-V2 models.
Figure~\ref{fig:training_budget} displays the smoothed generative training loss curves as a function of total compute (measured in FLOPs) for models of various scales.
Specifically, larger models need more computational resources to achieve the same loss value, but they also converge to lower loss points, which is consistent with observations in the field of large language models.

\begin{figure}[h]
    \centering
    \begin{subfigure}[b]{0.48\textwidth}
        \centering
        \includegraphics[width=\textwidth]{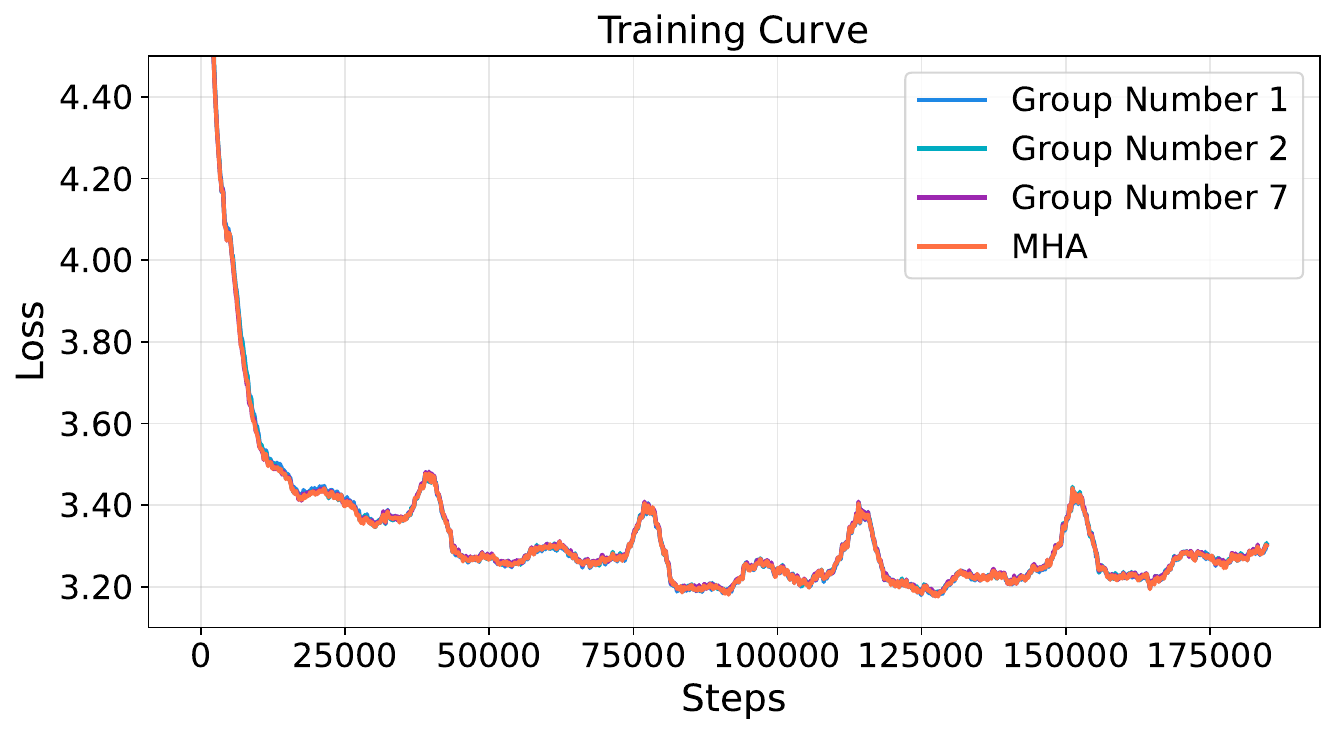}
        \caption{Key-value sharing.}
        \label{fig:kv_sharing_loss}
    \end{subfigure}
    \hfill
    \begin{subfigure}[b]{0.48\textwidth}
        \centering
        \includegraphics[width=\textwidth]{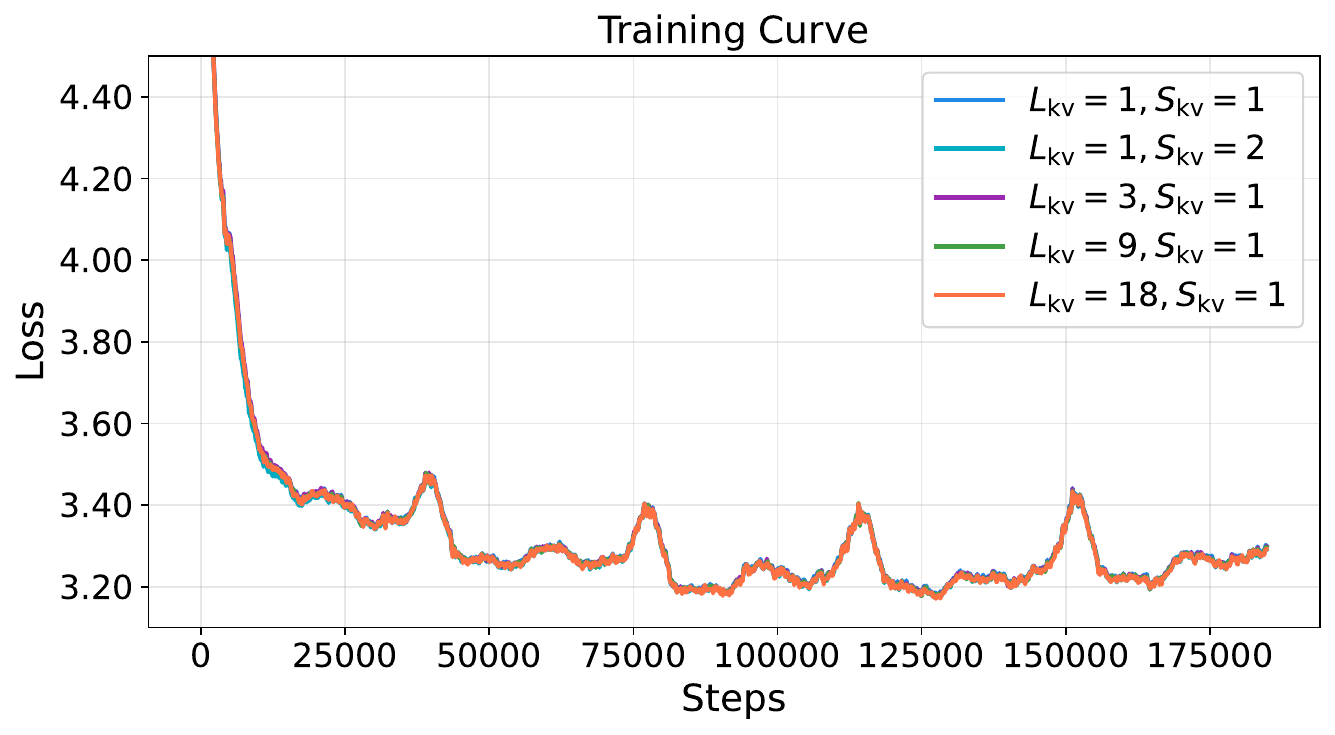}
        \caption{Grouped query attention.}
        \label{fig:gqa_loss}
    \end{subfigure}
    \caption{Training loss curves for different cross-attention configurations. Both key-value sharing (left) and grouped query attention (right) strategies show minimal impact on convergence loss while significantly enhancing computational efficiency.}
    \label{fig:combined_loss}
\end{figure}

\section{Online Performance with Caching Disabled}
\label{caching_disabled}
As mentioned in Section~\ref{sec:online}, our experimental group traffic is 5\%, with OneRec-V2 applied to 25\% of the degraded traffic within this group. For a more rigorous comparison, we allocate an additional 1\% experimental group with caching disabled (in this group, all traffic requests OneRec-V2). The performance is shown in Table~\ref{online_all_traffic}.

When all traffic requests OneRec-V2, we observe substantial improvements in key engagement metrics including watch time and user interaction indicators. Specifically, interaction metrics such as likes, follows, comments, and forwards demonstrate remarkable gains ranging from 9.6\% to 29.2\% across different platforms. However, certain ecosystem-level metrics present concerning trends. Notably, cold-start video views experience significant degradation (44.7\% and 36.7\% decline for Kuaishou and Kuaishou Lite respectively), while cluster density increases substantially (11.7\% and 7.9\%). This presents a critical challenge that requires careful consideration in our future directions.

\begin{table*}[t]
    \centering
    \caption{The relative improvement of OneRec-V2 compared to OneRec-V1 (in this group, all traffic requests OneRec-V2).}
    \renewcommand{\arraystretch}{1.2} 
    \setlength{\tabcolsep}{20pt}      
    \begin{tabular}{p{2.5cm}|l|c}
    \toprule
    \textbf{Scenarios} & \textbf{Online Metrics} & \textbf{OneRec-V2} \\
    \hline
    \multirow{9}{*}{Kuaishou} & App Stay Time & +0.405\% \\
    & Watch Time & +0.513\% \\
    & Video View & +0.938\% \\
    & Like & +15.024\% \\
    & Follow & +15.755\% \\
    & Comment & +29.249\% \\
    & Collect & +9.640\% \\
    & Forward & +24.741\% \\
    & \textbf{Cold-Start Video View} & \textbf{-44.704\%} \\
    & \textbf{Cluster Density} & \textbf{+11.692\%} \\
    
    \hline
    \multirow{9}{*}{Kuaishou Lite} & App Stay Time & +0.958\% \\
    & Watch Time & +2.456\% \\
    & Video View & -1.121\% \\
    & Like & +12.783\% \\
    & Follow & +21.376\% \\
    & Comment & +16.975\% \\
    & Collect & +12.886\% \\
    & Forward & +30.957\% \\
    & \textbf{Cold-Start Video View} & \textbf{-36.730\%} \\
    & \textbf{Cluster Density} & \textbf{+7.933\%} \\
    \bottomrule
    \end{tabular}
    \label{online_all_traffic}
\end{table*}

\end{document}